\documentclass[preprint2]{aastex} 

\usepackage{graphicx}

\slugcomment{For submission to Astrophys. J.}

\shorttitle{Small Inner Companions}
\shortauthors{Van Laerhoven, C. and Greenberg, R.}

\begin{document}

\title{Small Inner Companions of Warm Jupiters: \\ Lifetimes and Legacies}

\author{Christa Van Laerhoven}
\affil{Department of Planetary Science, 1629 E University Blvd, Tucson, AZ 85721}
\email{cvl@lpl.arizona.edu}

\and

\author{Richard Greenberg}
\affil{Department of Planetary Science, 1629 E University Blvd, Tucson, AZ 85721}

\begin{abstract}

Although warm jupiters are generally too far from their stars for tides to be important, the presence of an inner planetary companion to a warm jupiter can result in tidal evolution of the system. Insight into the process and its effects comes form classical secular theory of planetary perturbations. The lifetime of the inner planet may be shorter than the age of the system, because the warm jupiter maintains its eccentricity and hence promotes tidal migration into the star. Thus a warm jupiter observed to be alone in its system might have previously cleared away any interior planets. Before its demise, even if an inner planet is of terrestrial scale, it may promote damping of the warm jupiter's eccentricity. Thus any inferences of the initial orbit of an observed warm jupiter must include the possibility of a greater initial eccentricity than would be estimated by assuming it had always been alone. Tidal evolution involving multiple planets also enhances the internal heating of the planets, which readily exceeds that of stellar radiation for the inner planet, and may be great enough to affect the internal structure of warm jupiters.  Secular theory gives insight into the tidal processes, providing, among other things, a way to constrain eccentricities of transiting planets based on estimates of the tidal parameter $Q$.

\end{abstract}



\section{Introduction} \label{intro}

The discovery of vast numbers of extrasolar planets over the past two decades has revealed that many planets lie much closer to their stars than the planets in our solar system (e.g. \citealt{Bar13}). The main methods of discovery, radial velocity and transit observations, tend to favor close-in planets. Many are so close to their star that tides must have played significant roles in their orbital evolution, especially after the planet formation process was complete and the circumstellar nebula and disk had dissipated (e.g. \citealt{Jac08a}, \citealt{Jac08b}, \citealt{Mat10}). Tides tend to circularize orbits, reduce their semi-major axes, and heat the planets, so the current orbital distributions provide important constraints on physical properties, long-term histories, and past or present habitability. 

The current distribution of semi-major axes must reflect the process of removal of close-in planets as they spiral down into the star. As a planet gets too close, its tidal evolution accelerates, like driftwood approaching a waterfall. Thus, understanding tides may help explain the character of the distribution of close-in planets. For example, \citet{Jac08a}, \citet{Mat10}, and others have studied this effect for single-planet systems. The distribution of eccentricities also records the circularizing effect of tides. As the first close-in planets were discovered, their substantial eccentricities seemed inconsistent with the short ``timescale'' for tidal damping (e.g. \citealt{Ras96}, \citealt{Tri00}, \citealt{Bod03}). These substantial eccentricities can be explained by taking into account the coeval tidal effect on semi-major axis: These planets have spent most of their lives farther from the star, where eccentricity-damping was weaker \citep{Jac08a}. Another factor might be that a still-undiscovered planet (or planets) farther from a star could help maintain the eccentricity of an inner planet by gravitational perturbations. 

Indeed, multi-planet systems have been found to be so numerous that it is reasonable to expect that even where only a single close-in planet has been detected, there is plausibly at least one additional planet farther out. Here we consider the possible effects of secular planet-planet interactions on the tidal evolution of planets in a multi-planet system. Certainly, if the outer planet has an eccentric orbit, secular perturbation theory requires that the inner orbit must also be eccentric. In any planetary system, gravitational secular interactions periodically exchange angular momentum among the planets. Thus the effect of tides on a planet's orbit will be different than it would have been if the other planets were absent. In fact, the orbits of all the planets in a system evolve due to tides acting on only one of them.

In this paper we will show several examples of how important it is to consider the multi-planet nature of systems when analyzing their tidal orbital evolution. We focus on interior companions to warm jupiters in Sections \ref{infall} and {damping}. By definition (e.g. \citealt{Ste11}) warm jupiters lie beyond $\sim0.07$ AU. At that distance, any Jupiter-scale planet around a solar-scale star would experience little direct tidal evolution. An outer planet can rapidly drive its inner companion into the star (Section~\ref{infall}), reducing the multiplicity of those systems. Moreover, outer planets, even those relatively far from the star (far enough so that they would have minimal direct tidal evolution), can have their eccentricity damped strongly through tidal damping on the innermost planet (Section~\ref{damping}). Thus, a planet may have had a higher eccentricity in the past even if direct tidal effects on that planet have been negligible. In Section \ref{constrain} we analyze an observed system as a contrasting example and demonstrate how we can place constraints on current eccentricities that ensure the system has been stable throughout its lifetime (Section~\ref{constrain}).

\section{Incorporation of Tides into Secular Theory}

\subsection{Classical Secular Theory} \label{seculartheory}

We use the analytical framework of classical secular theory (\citet{Mur99} and \citet{Bro61}) to investigate the response of multi-planet systems to tides. Analytic solutions allow broader generalizations than computer intensive, N-body numerical integrations, which can only apply to the specific system being simulated. On the other hand, the analytic approach of secular theory constrains us to low-to-moderate eccentricities because it ignores terms  in the disturbing function that are higher than second order in eccentricity. Also, by considering only secular terms, we are constrained to non-resonant systems. Fortunately, many (if not most) multi-planet systems inhabit this region of parameter space 
In any case, numerical and analytical approaches are complementary, offering greater precision and physical interpretability, respectively.

In order to facilitate a secular solution, each eccentricity is described as a vector whose magnitude is the scalar eccentricity $e$ and with direction the longitude of pericenter $\varpi$. The Cartesian coordinates of the eccentricity vector are given by $k = e \cos\varpi$ and $h = e \sin\varpi$. To second order in the eccentricities, the differential equations governing a planet's eccentricity behavior are linear:

\begin{eqnarray} 
\dot{k_i} &=& - \sum_j{A_{ij} h_j} \label{kdot} \\
\dot{h_i} &=& \sum_j{A_{ij} k_j} \label{hdot}
\end{eqnarray}

\noindent where the matrix elements $A_{ij}$ describe the perturbing gravitational potential. (We order planets from innermost to outermost.) If only mutual gravitational perturbations are considered, the matrix $A$ depends only on the masses and semi-major axes of the system. Any additional effects on the precession of planet $p$ add to the magnitude of $A_{pp}$. Generally, the most significant such effect is the contribution of General Relativity (GR) to $A_{11}$, though the non-spherical figures of the planets and star can contribute as well \citep{Rag09}. In this paper, we include GR and precession caused by the planets' equilibrium tidal bulge. The solution to this set of linear differential equations (Eqn \ref{kdot} and \ref{hdot}) is a sum of eigenmodes:

\begin{eqnarray}
k_p = \sum_m{e_{mp} \cos(g_m t + \delta_m)} \label{kvst} \\
h_p = \sum_m{e_{mp} \sin(g_m t + \delta_m)} \label{hvst}
\end{eqnarray}

\noindent where $g_m$ is the eigenfrequency and $\delta_m$ is the phase of eigenmode $m$. For a system of $N$ planets there are $N$ eigenmodes. The ordering of eigenmodes is arbitrary. For consistency we number the modes in order of decreasing $g_m$ (i.e. $g_1$ is greatest).  In accordance with Eqns \ref{kvst} and \ref{hvst}, Figure \ref{ONE} shows how the eigenmodes will combine to describe the periodic behavior of a planet's eccentricity in a two-planet system. The eccentricity of each planet $\vec{e}_p$ is the vector sum of the rotating eigenmode components $\vec{e}_{mp}$, and each $\vec{e}_{mp}$ rotates in ($k$,$h$) space at a rate given by its eigenfrequency. Thus the eccentricity varies, although the magnitudes of all the $e_{mp}$ are fixed.

For each mode, the ratios among the various $e_{mp}$ values are given by the eigenvectors. For example, in a two-planet system, $e_{m1}/e_{m2}$ is given by the eigenvector ($V_{m1}$,$V_{m2}$) as $e_{m1}/e_{m2}=V_{m1}/V_{m2}$. As defined here, the eigenvectors are normalized such that $\sum V_{mp}^2 = 1$ where the summation is over all planets $p$. All $V_{mp}$ are functions of the coefficients $A_{ij}$ from Eqns \ref{kdot} and \ref{hdot}. Given the eigenvectors, each coefficient $e_{mp}$ in the solution (Eqns \ref{kvst} and \ref{hvst}) is the product of the amplitude $E_m$ of eigenmode $m$ and the normalized eigenvector component $V_{mp}$ for that mode and planet:

\begin{equation}
e_{mp} = E_m V_{mp} \label{emp}
\end{equation}

\noindent where the $E_m$ values are derived from the initial conditions (specifically the initial values of the eccentricities and pericenter longitudes of all the planets).

\subsection{Tidal Evolution Model} \label{tides}

To incorporate the effect of tides, we use the formulation presented by \citet{Gol66} and \citet{Kau68} which aggregated results from various sources (e.g. \citealt{Jef61}).
That model is based on a constant tidal dissipation parameter $Q$ that is independent of frequency, i.e. the tidal response has a constant phase lag for all Fourier components. Whether that assumption is a valid representation of the response of a real planet is uncertain, because tidal dissipation is an interplay of complex and only partially understood physical processes involving the friction-caused delay between the tidal strain and stress and self-gravity, along with turbulence, friction between layers, etc. Moreover, given these uncertainties, the standard practice of applying a lag to each Fourier component and summing the results may not always apply \citep{Gre09}, although for linear rheologies, such as the Maxwell rheology adopted by \citet{Dar79}, it is of course valid.

Because our approach retains only low-order terms in the tidal disturbing function, it is insensitive to high frequency effects, and thus is relatively independent of uncertainty in the frequency dependence of tidal responses. Also, as we retain terms in the gravitational disturbing function only through 2nd order in $e$, it is appropriate to go only to 2nd order in eccentricity in the tidal model as well. So, while other mathematical models (such the constant time lag model of \citealt{Hut81}) allow for expansion to higher order in eccentricity, those high-order terms may be less physically meaningful than their apparent precision may suggest and use of those higher order terms here would cause a mismatch in precision between the tidal model and the orbital dynamical model.

According to the formulation by Kaula and by Goldreich and Soter, for planets with orbits shorter than the spin period of the star,

\begin{equation} \label{adot}
\frac{1}{a} \frac{da}{dt} = - \left( \frac{e^2}{\tau_p} + \frac{1}{\tau_s}( 1 + \frac{57}{4}e^2) \right)
\end{equation}

\begin{equation} \label{edot}
\frac{1}{e} \frac{de}{dt} = - \left( \frac{1}{\tau_p} + \frac{25}{8} \frac{1}{\tau_s} \right)
\end{equation}

\noindent where

\begin{equation} \label{taup}
\frac{1}{\tau_p} = \frac{63}{4} \sqrt{GM_s^3} \frac{R_p^5}{Q_p m_p} a^{-13/2}
\end{equation}

\begin{equation} \label{taus}
\frac{1}{\tau_s} = \frac{9}{2} \sqrt{\frac{G}{M_s}} \frac{R_s^5 m_p}{Q_s} a^{-13/2}
\end{equation}

\noindent and $M_s$ and $R_s$ are the mass and radius of the star, $m_p$ and $R_p$ are the mass and radius of planet $p$, and $Q_s$ and $Q_p$ are the tidal dissipation parameter for the star and planet $p$ respectively (thus $Q_1$ is the tidal dissipation for the innermost planet, etc). Here, what we call $Q$ is actually $Q'$ of \citet{Gol66} where $Q' = 2 Q / 3 k_2$. Terms involving $\tau_p$ result from tides on the planet, and those involving $\tau_s$ result from tides on the star.

We can compare the relative strengths of planetary and stellar tides by taking the ratios of the terms due to each in Eqns \ref{adot} and \ref{edot}:

\begin{equation} \label{adotratio}
\frac{(da/dt)_{planetary}}{(da/dt)_{stellar}}
= \left( \frac{e^2}{1+(57/4)e^2} \right) \frac{\tau_s}{\tau_p}
\approx e^2 \frac{\tau_s}{\tau_p}
\end{equation}

\begin{equation} \label{edotratio}
\frac{(de/dt)_{planetary}}{(de/dt)_{stellar}} = \frac{8}{25} \frac{\tau_s}{\tau_p}
\end{equation}

\noindent where 
\begin{equation} \label{tauratio}
\frac{\tau_s}{\tau_p} = \frac{7}{2} \left(\frac{M_s}{m_p}\right)^2 \left(\frac{R_p}{R_s}\right)^5
\end{equation}

\noindent In this paper we use $Q_s \sim10^{6.5}$ \citep{Jac08a} and $Q_p\sim10^{5.5}$ for Jovian planets (\citealt{Jac08a}, \citealt{Yod81}). For rocky planets, $Q_p$ is $\sim10^3$ or smaller (\citealt{Jac08c} and references therein). For main sequence stars the combination of masses, radii and $Q$ in Eqn \ref{edotratio} is large, meaning that planetary tides will dominate the eccentricity evolution. For semi-major axis evolution, the ratio of contributions from planetary and stellar tides (Eqn \ref{adotratio}) is proportional to the same combination of masses, radii and $Q$ as in Eqn \ref{adotratio}, but multiplied by $e^2$ (Eqn \ref{adotratio}). Thus, $e$ must be very low for stellar tides to dominate the planet's semi-major axis evolution. Figure \ref{plaVSste} shows $\tau_s/\tau_p$ as a function of stellar mass ($M_s$) for planets of various sizes and tidal dissipation factors.

In addition to causing orbital evolution, tides generate heat in the periodically distorted body. The surface heat flux (in $W/m^2$ that results is:

\begin{equation} \label{heat}
h = \frac{63}{16 \pi} \frac{(GM_s)^{3/2} M_s R_p^3 e_p^2}{Q_p} a^{-15/2}
\end{equation}

\noindent This result follows from considering the energy budget of the system. Energy dissipated in the planet can only come from its orbit, not from its rotation, because the planet would have spun-down much earlier. Furthermore, the energy loss can only come from the planet's changing semi-major axis, because orbital energy depends only on $a$, not $e$. Thus, by using the relationship between orbital energy and semi-major axis, Eqn \ref{heat} follows from Eqn \ref{adot} above \citep{Jac08a}. The heating can be significant. For example, tidal heating likely plays a role in generating the radius anomalies of some hot jupiters (\citealt{Jac08b}, \citealt{Ibg11}, \citealt{Bod01}, \citealt{Mil09}).

\subsection{Combination of Tides and Secular Theory} \label{secularplustides}

Secular interactions exchange angular momentum, but not energy, among the planets. Consequently, each planet's semi-major axis changes at a rate given by Eqn \ref{adot} above, identical to what it would if the planet were a single planet. However, the tidal eccentricity damping of one planet will be shared among the system's eigenmodes. Thus all the planets' eccentricities will evolve differently than if they were single (e.g. \citealt{VLGrCM}). Moreover, because Eqn \ref{adot} for $\dot{a}$ depends on $e$, the long-term evolution of the semi-major axes of all planets will be affected by the combination of tides on one planet and the secular interactions.

Tidal evolution occurs slowly compared with the periodic secular changes in eccentricity (i.e. the process is adiabatic). Thus we can average the tidally driven change in the semi-major axes and eigenmode amplitudes $E_m$ over the secular cycle. This averaging can be done analytically. \citet{Gre11} derived formulas for $\dot{a}_p$ and $\dot{e}_{mp}$ (Eqns 16 and 19a-19d in that paper) for evolution due to tides on the inner planet. In Appendix \ref{appA} we extend that analysis to derive formulas for $\dot{a}_p$ and $\dot{E}_{m}$, including tides on both planets in a two-planet system and tides on the star. In order to study the evolution of a system, one can integrate those equations over time numerically.

The long-term evolution of a two-planet system is greatly simplified if one mode damps much faster than the other. Under what conditions does that occur? One can often ignore dissipation in the second planet because of the strong dependence of tidal effects on distance, so to address this question we consider only the effects of tides on the inner planet in the following argument (though we include tides on both planets in all our integrations). For a damping process acting on the inner planet's eccentricity, according to Appendix \ref{appA}, the ratio $D$ of the resulting damping of the two eigenmode amplitudes is given by

\begin{equation} \label{reldamp}
D \equiv \frac{\delta E_1 / E_1}{\delta E_2 / E_2} = \frac{- V_{11} V_{22}}{V_{12} V_{21}}
\end{equation}

\noindent (note that $V_{11}$ is negative so $D$ is positive, see Appendix \ref{appA}). The larger the value of $D$, the more of the eccentricity damping is partitioned to mode 1 (by definition the mode with the greater eigenfrequency). For a case where planet-planet interactions dominate (and other effects like GR are negligible), $D$ is a function of only $a_1/a_2$ and $m_1/m_2$, as shown in Figure \ref{DAMP1}. Also, for any given $a_1/a_2$, $D$ decreases monotonically with $m_1/m_2$ (Figure \ref{DAMP1}). For some choices of $a_1/a_2$ and $m_1/m_2$, $D\sim1$, i.e. both eigenmodes damp on comparable timescales. For examples of such systems see \citet{VLGrCM}, \citet{VLGrKep}, and Section \ref{constrain} below. For many cases of interest $m_1/m_2<1$, in which case $D>1$ for any choice of $a_1/a_2$, i.e. eigenmode 1 will always damp faster than eigenmode 2.

So, under what conditions will mode 1 damp \textit{much} faster than mode 2? If the planets are comparably massed (if, for example, they were both Jovian or both terrestrial) and $a_1/a_2$ is less than $\sim0.2$, then $D>10$. Also, if the inner planet is much smaller than the outer planet (i.e. the inner planet is terrestrial and the outer planet is Jovian), then for any $a_1/a_2$ less than $\sim0.95$, $D$ would be $>10$. These situations would thus have the amplitude of mode 1 ($E_1$) damp much faster than that of mode 2 ($E_2$).

Because effects that act only on the precession rates (e.g. GR) contribute only to the diagonal elements of the matrix $A$, including these effects would increase $V_{11}$ and $V_{22}$ (Eqn \ref{F1} and \ref{F2}), thus increasing $D$.

\section{Rapid In-fall} \label{infall}

For a single planet, Eqns \ref{adot} and \ref{edot} show that its $\dot{e}/e$ due to planetary tides is always faster than $\dot{a}/a$ ($|\dot{a}/a| < |\dot{e}/e|$). Thus, the planet's tidal eccentricity damping will happen quickly relative to its semi-major axis migration. When the planet's eccentricity becomes so small that tides on the planet are negligible, semi-major axis evolution will still continue due to tides raised on the star, but only slowly. Such a planet can survive for a very long time before being destroyed as it reaches the star.

In contrast, in a multi-planet system the damping of one or more of the eigenmode amplitudes ($\dot{E}_m/E_m$) can be slower than $\dot{a}/a$ for either of the planets. In other words, a tidally worked planet can have its eccentricity remain significant much longer than if it were the only planet. But its semi-major axis will decrease more quickly because $\dot{a}$ due to planetary tides is proportional to $e^2$ (Eqn \ref{adot}). This planet will therefore have a shorter lifetime than it would if it were a single planet. In other words, the planet can reach the Roche zone of the star even with minimal contribution from tides on the star.

\subsection{An Example: HAT-P-15 b}

Consider the tidal evolution of  planet HAT-P-15b, with current mass $m = 2 M_{Jup}$, radius $R = 1.072 R_{Jup}$, $a = 0.096 AU$, and $e = 0.19$ orbiting a G5 star of mass $M_s= 1.95 M_{Sun}$ and age $\sim7$ Gyr \citep{Kov10}. We adopt for the uncertain tidal dissipation parameters the plausible values $Q_p =10^{5.5}$ and $Q_s = 10^{6.5}$ (see Section~\ref{tides}). Assuming the planet is and has always been single, integration of Eqs. \ref{adot} and \ref{edot} back in time yields the evolution shown by the dashed lines in the right-hand panels of Figure \ref{HATa}. The orbital changes are modest and nearly linear with time, with $a$ changing by less than 1\% and $e$ reduced from $0.23$ to $0.19$.  The tidal heating rate correspondingly decreased by about 30\%.

Suppose Hat-P-15b was originally (7 Gyr ago) accompanied by an interior rocky planet, of mass $m_1 = 5 M_{Earth}$ and $Q_1 = 10^3$, with $a_1 = 0.045 AU$ and $e_1 = 0.135$. Throughout this paper, for the rocky planets we use $R_p = (m_p/M_{Earth})^{1/2.06} R_{Earth}$ \citep{Lis11}. If such a rocky planet were unaffected by the outer planet, integrating Eqns \ref{adot} and \ref{edot} yields the orbital evolution shown by the dashed lines in the left-hand panels of Figure \ref{HATa}. The eccentricity $e_1$ damps away so quickly (on a timescale of 20 Myr) that tidal dissipation nearly stops and $a_1$ remains essentially constant.

Next we consider the evolution of the two-planet system (the solid curves in Figure \ref{HATa}). The initial conditions for the inner planet (at 7 Gyr ago) are the same as defined above; for the outer planet, we have selected initial conditions that lead to the current observed orbit of HAT-P-15 b. For this case, Eqn \ref{reldamp} (along with Figure \ref{DAMP1}) shows that mode 1 dominates the damping of the inner planet; Correspondingly this mode damps at a rate comparable to the damping of $e_1$ for the case where the outer planet is ignored, i.e. mode 1 damps away in $\sim20$ million years. For studying the evolution over billions of years, we assume that mode 1 has damped away by the start of this evaluation, and set the amplitude $E_1$ of this mode to zero. With only one eigenmode, there is no oscillation of either planet's eccentricity over secular timescales, so $e_1 = e_{21}$ and $e_2 = e_{22}$. The evolution is shown by the solid curves in Figure \ref{HATa}.

For the inner planet (Figure \ref{HATa} (left)), starting from the initial values, both $a_1$ and $e_1$ decrease gradually over the course of the first 400 Myr.  The change in $a_1$ is due to the combined effect of tides raised on the inner planet on the star and on the star by the planet. The contributions of these two effects over the course of the evolution are shown in Figure \ref{HATb} (top). During these first 400 Myr, tides raised on the planet are dominant in reducing $a_1$.

At the same time, tides cause $e_1$ to decrease. With secular mode 1 assumed to be negligible, the tidal damping only affects the amplitude $E_2$ of mode 2. $E_2$ is damped by tides raised on both planets and on the star. Figure \ref{HATb} (middle) shows the effect of tides raised on the star and on both planets. The latter is dominated by inner-planet tides. During the first 100 Myr, $E_2$ damps at a rate $\sim10$\%/Gyr.  However, the evolution shown in Figure \ref{HATa} shows $e_1$ damping much faster, changing by $>10$\% during the 100 Myr.  A similar discrepancy continues for at least the first 400 Myr. 

(Though we have set the amplitude of $E_1$ to zero, we show $\dot{E}_1/E_1$ in Figure \ref{HATb} (middle). As can be seen in this figure, the damping rate of $E_1$ is indeed much much faster than that of $E_2$, as predicted from looking at the behavior of $D$ (Eqn \ref{reldamp}, Figure \ref{DAMP1}).)

Why does $e_1$ decrease so quickly, even though the amplitude of the eigenmode does not? Recall that $e_1$ also depends on the eigenvector, which describes how much of mode 2 is shared with the inner planet:  $e_1 = E_2 V_{21}$.  Although the tidal damping of the amplitude $E_2$ is very slow, $V_{21}$ changes much more quickly as shown in Figure \ref{HATb} (middle).  In fact, its rate of change of about $-10$\% per 100 Myr is close to that of $e_1$.  Thus the decrease of $e_1$ is not due to direct tidal damping of the eccentricity, but rather to the change in $a_1$ which results in a change in the partitioning of the magnitude of mode 2 between the planets.  As $a_1$ decreases, the strength of the secular coupling between the planets diminishes, until mode 2 predominantly affects the outer planet and has little effect on the inner one.  Consequently $e_1$ decreases rapidly.

During this period, the decrease in $a_1$ accelerates, because of the strong dependence of $\dot{a}_1$ on $a_1$ (Eqn \ref{taup}).  At the same time, the change in $e_1$ accelerates as well, because of the uncoupling of the secular interaction. Eventually, after about 430 Myr, $e_1$ is so small that the change in $a_1$ begins to decelerate (Figure \ref{HATb} top), although the planet does continue to migrate inward to the planet. After 440 Myr, the planet is close enough to the star that the contribution of tides on the star to $\dot{a}_1$ (which is independent of eccentricity) begins to become significant, and the inward migration accelerates again.  Just after about 460 Myr from the start, the planet reaches the Roche limit where it would be tidally disrupted.

During the entire lifetime of the inner planet, the decrease of $e_1$ tracks the change in $a_1$, as the planets become decoupled and a decreasing share of eigenmode 2 goes to the inner planet. This decoupling is illustrated in Figure \ref{HATb} (bottom), where the component that affects the inner planet ($V_{21}$) decreases relative to the share that goes to the outer planet.

The outer planet's eccentricity also damps during this period as the damping of the second eigenmode affects both planets.  Note that if the eigenvector ($V_{12}$,$V_{22}$) remained constant over the 460 Myr, $e_2$ would damp in proportion to $e_1$.  However, Figure \ref{HATa} (right) shows that $e_2$ drops by only $\sim10$\% during this period (compared with nearly 100\% for $e_1$), because of the change in the eigenvector as the planets becomes less coupled.  Just as the change in the eigenvector accelerates the decrease of $e_1$, it slows the decrease of $e_2$, which is actually dominated by the damping of the amplitude $E_2$ of eigenmode 2. After the inner planet is destroyed, the tidal evolution of the orbit of the outer planet continues at a modest rate until the current orbit of HAT-P-15 b is reached.

This example demonstrates how the presence of an outer planet can accelerate the tidal migration and subsequent demise of an inner planet.  The fact that a ``warm jupiter'' like HAT-P-15 b currently displays no evidence of any inner planets does not imply that it formed in a system with no inner planets.

\subsection{Parameters that Speed an Inner Planet's Demise}

All planets that have an outer companion have a shorter lifetime than they would as singles. However, this effect is more dramatic for some parameters than for others. For a given outer planet, the larger $a_1$ is, the greater is $V_{21}/V_{22}$ and thus $e_1$. This greater $e_1$ increases tidal dissipation tending to reduce the planet's lifetime. On the other hand, tides also depend strongly on $a_1$ (Eqn \ref{taup} and \ref{taus}). The dependence of $\tau_s$ and $\tau_p$ on $a_1$ (Eqn \ref{taus} and \ref{taup}) is much stronger than that of $e_1$. So, the smaller $a_1$ is, the shorter the lifetime. But, that dependence applies whether the planet is single or has an outer companion. If we compare the single-planet case to the multi-planet case, the latter always has a shorter relative lifetime. The larger $a_1$ is, the greater this difference because of the increased $e_1$ driven by the outer planet.

\subsection{Tidal Heating} \label{deathheat}

A potentially important consequence of a planet's eccentricity history is the effect on tidal heating, because the heat generation is proportional to $e^2$ (Eqn \ref{heat}). The bottom panels of Figure \ref{HATa} show the tidal heat generated within HAT-P-15 b and its possible inner companion. Because the inner rocky planet's eccentricity is sustained by the outer planet, its level of tidal heating is not turned off after 100 Myr as it would be if the planet were single. Moreover, due to the strong $a$-dependence of $h$ (Eqn \ref{heat}), as the rocky planet's semi-major axis decreases, the level of tidal heat increases. As the planet approaches the star, tides dominate the planet's heat budget.  Tidal heating rises sooner and faster than the stellar insolation (or ``instellation''), exceeding the latter by about a factor of two for tens of millions of years, until just before the planet's demise. Only as the inner planet nears the star does its eccentricity decrease enough for the tidal heating to decrease to near the instellation rate.  The surface flux of internal tidal heat is orders of magnitude greater than that of Io, so this 5 Earth mass planet would undergo violent geophysical transformation during the last 20\% of its lifetime.

As shown in Figure \ref{HATa}, the outer planet HAT-P-15 b also experiences a boost in tidal heat generation. While the inner planet remains in orbit, the outer planet's heating rate is boosted by 20\% to a maximum over 20 $W/m^2$, equivalent to $10^{18} W$ total dissipation for $\sim$ 1 Gyr. 

Tidal heating has been proposed as a possible energy source to explain some hot jupiters' radius anomalies (\citealt{Bod01}, \citealt{Jac08b}, \citealt{Mil09}). Heating at a rate of $\sim5\times10^{19} W$ within the past $\sim1$ Gyr might be adequate to have yielded the puffed-up radii \citep{Bur07} and some hot jupiters might have been puffed up thanks to tidal heating \citep{Jac08b}. In general, a given warm jupiter (such as HAT-P-15 b at 0.1 AU) would have nowhere near enough tidal heat to explain a radius anomaly (e.g. \citealt{Mil09}). However, with our hypothetical, now-defunct inner planet, the warm jupiter HAT-P-15 b could have experienced considerably more heating during its history than if it had always been alone. While this value is very likely too small to have affected the planet's radius, it may have been great enough to have affected the internal physical processes of such a planet early in its history. More to the point of the current study, depending on the parameters of any particular system, a warm jupiter with a small inner companion that is either now lost to tidal evolution or too small to be detected, may have experienced heating great enough and recent enough to have detectable physical consequences.  A systematic search for such conditions is left for future work. The consequences of multi-planet interactions on another warm jupiter's tidal heating history are discussed further in the Section \ref{dampingheat}.

\subsection{Current Companions to HAT-P-15 b} \label{HATcompanions}

In the above example, the inner planet's lifetime was very short, only a tenth of the estimated age of the system.  However, a less massive inner planet might have survived the tidal evolution, even given its enhanced eccentricity driven by the outer planet, and still be in orbit around HAT-P-15. For example, by integrating backwards in time, we find that an Earth-mass planet with $Q=1000$ could have started far enough from the outer planet to be stable, and also well inside the 3:1 resonance, and still be in orbit at 0.03AU after about 6 Gyr, comparable to the estimated age of the system. Other examples are shown in Figure 6 for various masses and current semimajor axes, all assuming $Q=1000$ for the inner planet.

In fact, inner planets could still be in orbit having survived even longer then the durations shown in Figure \ref{HATmany}.  If a planet formed outside the 3:1 resonance, but still far enough from HAT-P-15 b that its initial orbit were stable, tidal evolution would have carried it into the resonance. It could then have jumped through the resonance. Depending on the specifics of the resonance jump, particularly how it affected the eccentricities, this planet could have then continued migration and still be present in the system. The duration of evolution shown in Figure \ref{HATmany} only refers to the time after the resonance passage.

Thus the substantial mass and eccentricity of HAT-P-15 b cannot be assumed to have led to tidal removal of all inner planets.  It is possible that the system still includes an inner planet of about one Earth mass, or even larger if the system is younger than its estimated age. Of course, this result depends on the assumed value of $Q$.  A truly Earth-like planet would have a much smaller $Q$ and thus a shorter lifetime, but there is no reason to expect a planet so close to its star to have geophysical properties so similar to those of Earth.

\section{Damping the Eccentricity of Outer Planets} \label{damping}

Next we consider how even a modestly sized inner planet can share the effects of tides in ways that significantly affect the outer planet, through the mechanism of secular interactions. Thanks to tides, hot jupiters' orbits may be more nearly circular now than in the past (e.g. \citealt{Jac08a}, \citeyear{Jac08b}). Warm jupiters, on the other hand, are generally too far out to have experienced significant direct tidal damping. However, even such warm jupiters may have had their eccentricities reduced from much larger values if a modest-sized inner planet had been present for part of the life of the system.

Due to the secular coupling of the planets, the direct tidal action on the inner planet also would be propagated to the outer warm jupiter. As we have seen, eccentricity damping is shared among all the eigenmodes, so other planets than the one being directly affected by tides also experience eccentricity damping. Thus, in general, planets exterior to an inner, tidally-worked planet may well experience significant eccentricity evolution even if they themselves are too far from their star to be directly affected much by tides. Even if tides have already driven the inner planet in to the star, the remaining warm jupiter would have had a history that involved a much more eccentric orbit than we might imagine if we assume it was always single. Such a history can be illustrated with the following example.

\subsection{An Example: HD 130322 b} \label{dampingex}

Consider the warm jupiter HD 130322 b with mass $1.02 M_{Jup}$, current $a=0.088$ AU and $e=0.044$, and estimated age of $\sim$0.35 Gyr (\citealt{Udr00}, \citealt{Saf10}). Its radius is unknown, but we assume $R= 1 R_{Jup}$. It orbits a star of mass 0.79 $M_{Solar}$ and radius 0.83 $R_{Solar}$. If HD 130322 b has been single since its placement close to its host star, integrating Eqns \ref{adot} and \ref{edot} with $Q$ values of $10^{5.5}$ for the planet and $10^{6.5}$ for the star, shows that this planet would have experienced essentially no eccentricity or semi-major axis evolution, as shown by the dashed lines on the right side of Figure \ref{HD10a}.

Now suppose HD 130322 b initially had an inner companion with $m_1=10 M_{Earth}$, $Q_1=500$. Now an entirely different orbital history can lead to the currently observed system, with the same orbit of planet b, and the hypothetical inner planet having fallen into the star (Figure \ref{HD10a}, solid lines). Here we have selected initial orbits that lead to this current configuration, and set the amplitude of mode 1 to zero because it would be very short lived.

In this scenario, the semi-major axis evolution of HD 130322 b (Figure \ref{HD10a}, top right) is nearly indistinguishable from that of the single-planet case, that is, there is almost no semi-major axis change. The eccentricity evolution, however, is quite different. Indeed, the initial eccentricity of HD 130322 b with this modest-sized companion would have been 0.185, more than four times its current value. Implications for tidal heating (bottom of Figure \ref{HD10a}) are discussed in Section~\ref{dampingheat}.

This dramatic difference in evolution in the multi-planet case versus the single-planet case is due to tides acting on the inner companion planet. In contrast to the example presented in Section \ref{infall}, the decrease in $e_1$ in this situation is not primarily controlled by the changing eigenvector $V_{21}$ (which depends on evolution of the planets' semi-major axes). Instead, the eccentricity of the inner planet decreases due to decreasing $E_2$ (a consequence of eccentricity damping), as well as the decreasing $V_{21}$. Indeed, $\dot{E}_2/E_2$ is greater than $\dot{V}_{21}/V_{21}$ for most of the inner planet's life, as shown in Figure \ref{HD10b} (middle). Only for the first 80 Myr and last 20 Myr does $\dot{V}_{21}/V_{21}$ overcome $\dot{E}_2/E_2$.

If the outer planet's eccentricity change were only due to the change of the eigenvector then its eccentricity would increase. As the inner planet's semi-major axis decreases, $V_{21}$ also decreases. Because the eigenvectors are normalized, this means $V_{22}$ increases. However, $E_2$ is decreasing as shown in Figure \ref{HD10b} (middle). Like $V_{21}$, the rate of change of $V_{22}$ is weak, too weak to overcome the effect of changing $E_2$. Hence  the outer planet's eccentricity decreases along with that of the inner planet.

\subsection{A Smaller Inner Planet} \label{dampingsize}

The amount by which the eccentricity of the outer planet changes is affected by the mass and size of the inner planet. For tides raised on the star, $\tau_s \propto m_p^{-1}$ so a more massive inner planet will tend to have faster evolution, a consequence of its greater ability to perturb the shape of the star. For tides raised on a planet $\tau_p \propto m_p/R_p^5$, so the bigger the planet, the faster the orbits evolve, unless the planet is extraordinarily dense.

We can see how this manifests for HD 130322 by next considering a smaller inner planet with $m_1=5 M_{Earth}$, $R_1=2.2 R_{Earth}$ and still with $Q_1=500$. In this case HD 130322 b's initial eccentricity could have been reduced from 0.65 before reaching its current value, simply by the presence of this now-lost inner planet.

Bear in mind that we selected an initial orbit for this inner planet that would maximize its lifetime, to maximize its effect on damping the eccentricity of the outer planet, before it itself disappears from the system. If the system's age were actually 0.75 Gyr, the inner planet could have started at 0.035 AU and still have had time to fall into the star recently. With that extra time, HD 130322 b's eccentricity could have been damped from an initial value of 0.138 to its current small value. Comparing this case to the case shown in Figure \ref{HD10a} (with a 10 $M_{Earth}$ inner planet), the inner planet goes though the same semi-major axis change and has a similar effect on the outer planet's eccentricity. Thus, a 5 $M_{Earth}$ inner planet could have been about as effective as a 10 $M_{Earth}$ one if it has the extra time.

\subsection{Parameters that Maximize an Inner Planet's Effect} \label{dampingparams}

While having an inner companion will always damp all the eigenmode amplitudes, some parameter space is more conducive for affecting the outer planet's eccentricity. To damp the outer planet's eccentricity at least one of $E_2$ and $V_{22}$ must decrease (and if only one of those is decreasing its effect must be dominant). As the inner planet approaches the star, the greater separation between planets will slowly uncouple their interaction, meaning $V_{11}$ and $V_{22}$ will increase. So, to damp the outer planet's eccentricity, $E_2$ must damp. Considering a case where the process is to run to completion (i.e. the inner planet has time to fall all the way from wherever it started down into the star), the effect can be maximized by having the rate $E_2$ decrease as fast as possible, while $a_1$ decreases as slowly as possible (to maximize the amount of time the inner planet survives to get worked by tides).

As discussed in Section \ref{secularplustides}, we can ascertain what $\dot{E}_2/E_2$ is compared to $\dot{E}_1/E_1$ through the ratio $D$ (Eqn \ref{reldamp} and Figure \ref{DAMP1}). As shown in Figure \ref{DAMP1}, for any given outer planet, larger values of $a_1$ and/or $m_1$ will result in stronger damping of $E_2$ compared to $E_1$. However, while a larger $a_1$ will raise the rate at which $E_2$ damps relative to the rate a which $E_1$ damps, it will also decrease the absolute speed of $E_2$ damping due to the dependence of $\tau_1$ ($\tau_p$ for $p=1$) on $a_1$. This means that damping of $E_2$ will take longer for a larger initial value of $a_1$.

As discussed in Section \ref{infall}, a planet in either a multi- or single-planet system will live longer if its initial $a_1$ is larger. Thus, a larger initial $a_1$ will ultimately lead to more damping of $E_2$, and thus $e_2$, if the process has time to run to completion. 

In addition to looking at what $a_1$ and $m_1$ will produce more damping on a given outer planet, we can also look at what role the inner planet's eccentricity plays. Both $\dot{E}_2$ and $\dot{a}_1$ depend on $e_1$, however $\dot{a}_1$ has a $e_{1}^2$ dependence (Eqn \ref{adot}), but $\dot{e}_1$ (Eqn \ref{edot}), and thus also $\dot{E}_2$ (see Appendix \ref{appA}), only depends on $e_1$ linearly. Thus, at lower eccentricities $\dot{a}_1$ will be comparatively weaker versus $\dot{e}_1$. Therefore, an outer planet's eccentricity will be more strongly affected at low eccentricities.

\subsection{Tidal Heating} \label{dampingheat}

The effect of an inner companion on the eccentricity of the outer planet leads to greater tidal heating of the outer planet than if that planet were alone. For HD 130322 b, the 10 Earth mass inner companion would for $\sim100$ Myr result in an order of magnitude greater tidal heating in the outer planet than if such an inner planet never existed (bottom right panel of Figure \ref{HD10a}). The 5 Earth mass planet results in about 3 times as much tidal heating of the outer planet compared to that outer planet being single. During this time the tidal dissipation rate is $\sim10 W/ m^2$, half the heating rate for the outer planet in our hypothetical HAT-P-15 system, and for only $\sim1/10$ the duration (c.f. Section \ref{deathheat}). Nevertheless, this dissipation rate (equivalent to $\sim5\times10^{18} W$) could conceivably have played a role in the remaining planet's geophysical history, perhaps affecting the current physical properties.

The sustained eccentricity of the inner planet results in great tidal heating of the inner planet. The 10 Earth mass inner planet experiences tidal heating comparable to the instellation for most of its lifetime. The 5 Earth mass planet receives only about an order of magnitude more heat from instellation than from tidal heat. (For comparison, Earth receives about $10^4$ times more heat from the Sun than from radiogenic sources.) As for HAT-P-15 the heating is thanks to the role of the outer one planet maintaining the inner planet's eccentricity, as well as the smaller values of $Q$ that we have assumed for these smaller hypothetical planets. Again, the tidal heat flux is orders of magnitude greater than that of Io, so major geophysical processing would be expected.

\subsection{Current Companions to HD 130322 b}

In Section \ref{HATcompanions} we showed that the HAT-P-15 system could still include an Earth-scale planet, even given the tidal migration toward the planet.  For HD 130322, the young estimated age of the system makes the survival of an inner planet possible for a much wider range of masses and current semimajor axes.

As shown in Figure \ref{HDmany}, any planet with a mass up to 10 Earth masses, or even more, could remain in orbit for much longer than the estimated 0.35 Gyr age of the system, assuming $Q=1000$. The overall pattern remains the same as for HAT-P-15 b: higher massed companions evolve faster than lower mass ones. Nevertheless for this system, tidal evolution cannot rule out the current presence of an additional massive planet interior to the orbit of the known giant planet.

\section{Constraining Eigenmode Amplitudes Given $Q$, or Vice-Versa} \label{constrain}

For planetary systems that currently have more than one planet, consideration of possible orbital histories can place constraints on tidal parameters. As shown in Sections \ref{infall} and \ref{damping}, where the inner planet experiences greater tidal evolution, the original orbits must have been much closer together. However, two planet systems are only stable if the planets come no closer than $\sim$2.5 Hill Radii from each other (\citealt{Gla93}, \citealt{Smi09}), which places constraints on tidal parameters.

Consider the case of KOI-543, an unconfirmed system of two planets orbiting a star of radius $0.75 R_{Sun}$, mass 0.85 $M_{Sun}$, and age $\approx0.5$ Gyr (\citeauthor{NASAExoArch}). The planets have radii 1.42 $R_{Earth}$ and 2.29 $R_{Earth}$, semi-major axes $a_1=0.040$ AU and $a_2=0.049$ AU, and periods 3.1 and 4.3 days, respectively. (Note that here we number the planets in order of increasing orbital period. These planets have the reverse order in the Kepler KOI catalog.) We adopt masses of $m_1=2.1 M_{Earth}$ and $m_2= 5.5 M_{Earth}$ based on the mass-radius relationship of \citet{Lis11}.

Because tides on both planets are important, and the planets have similar masses and tightly packed orbits (such that $V_{12}$ and $V_{21}$ are not small compared to $V_{11}$ and $V_{22}$), both eigenmodes damp on similar timescales. Therefore, we assume that the two eigenmodes have equal amplitude at present. Moreover, they damp much faster than the changes in semi-major axis.

Assuming for simplicity that both planets have the same value of $Q_p$, if these were rocky planets (with $Q$ less than order of 100), the tidal damping would be extremely fast. Even with $Q_p=1000$, the damping timescale is only about 100 Myr, as shown in Figure \ref{K5431000a}. For the evolution shown in that Figure, we assumed current eccentricity eigenmode amplitudes of $10^{-4}$. Figure \ref{K5431000b} (top) confirms that the two eigenmode amplitudes damp on very similar timescales. Figure \ref{K5431000b} (bottom) shows that the planets' minimum separation would have been only barely over the limit of stability (2.5 $R_{Hill}$) at 0.5 Gyr ago. Thus, given the estimated age of the system, unless the current eccentricities are extremely small, $Q_p$ cannot be less than 1000. Thus, if $Q_p = 1000$ the current mode amplitudes can be no higher than $10^{-4}$ unless some mechanism increased or sustained them.

For the case shown in Figure \ref{K5431000a} with both planets' $Q = 1000$, these planets experience extreme tidal heating. Even with no instellation, for $\sim100$ Myr the internally generated tidal heat flux was two orders of magnitude greater than that of Io. However, the tidal dissipation rate is quickly reduced as the tides also damp the eccentricities. At 0.5 Gyr ago, the heat fluxes of about a couple hundred $W/m^2$ would yield a surface temperature of $\sim 250$ K from the Stefan-Boltzmann Law. However, the instellation is much greater than the tidal heat flux, so the 250 K temperature would only be relevant to the night-side of the planets, and then only if the atmosphere was unable to distribute stellar heat globally. 

If both planets have $Q_p$ of $5\times10^4$, which is more like that of Saturn \citep{Gol66}, then their eccentricities would have changed little over the past 0.5 Gyr. With this $Q_p$, the current eccentricity mode amplitudes could still be as high as $\sim0.065$. The tidal heat flux for the inner and outer planet are $\sim12 W/m^2$ and $\sim5 W/m^2$ for the entire 0.5 Gyr. These heat fluxes are still somewhat greater than that of Io, but the assumed $Q$ of $5\times10^4$ implies the planets would be gassy, so this heat flux might not have the same consequences as it does for rocky Io.

Based on their radii it is likely that the inner planet is rocky and the outer one has a significant gas envelope (L. Rogers, private communication). Accordingly we consider a case  where the outer planet's $Q$ is $10^{5.5}$, while the inner one's is $1000$. If both eigenmode amplitudes ($E_1$ and $E_2$) currently have the same value, the largest they can be is $10^{-3}$. The evolution is shown in Figures \ref{K543mixeda} and \ref{K543mixedb}. The first eigenmode damps to the current value about three times faster than the second eigenmode (Figure \ref{K543mixedb}). Thus, unlike the first two cases (with equal $Q$ values for both planets) where the difference between maximum and minimum eccentricity changed on a timescale similar to the over-all damping of the planets' eccentricities, in this case the difference between maximum and minimum eccentricity changes more slowly than the mean eccentricity does. This results in somewhat higher tidal heat flux for the inner planet than in the 1st case (where both $Q = 1000$) because, while the maximum $e$ of this planet behaves similarly in both cases, the average $e^2$ for the inner planet is higher in this case.

To summarize this section, if both planets in this system have a $Q$ typical of a rocky planet, their eccentricities should be very near zero. Indeed, rocky planets typically have $Q$ values smaller than the $Q_p=1000$ used in the first case presented in this section. However, if both planets have a $Q$ more typical of a gaseous planet, then the planets could still have significant eccentricities. In the more realistic, 3rd case presented here (though again, we've used a $Q$ for the inner planet that is larger than expected for a rocky body), we again see that the current mode amplitudes (and thus eccentricities) must be small ($< \sim 10^{-3}$) for the system to have been stable.

\section{Discussion and Conclusions}

It has long been recognized that the tidal damping of an inner planet's eccentricity can be counteracted by the effect of another planet.  However, as demonstrated here, the interplay of gravitational perturbations and tidal evolution is more complicated than that, even where the perturbations are controlled by secular interactions alone, without the effects of resonances. A single warm jupiter would be too far from its parent star to have been directly affected by tides.  However, with the presence of an additional inner planet, even one that no longer exists or is too small to be observed, the warm jupiter's tidal evolution and geophysical processing could be significant.

We have seen how the reduction in the eccentricity of a warm jupiter like HAT-P 15 b to its current value could have been doubled over its lifetime given the presence of a terrestrial-scale inner planet. But the small culprit would have vanished long ago, driven into the star by the combined effects of tides and planetary perturbations. HD 130322 b alone would have undergone negligible tidal evolution. On the other hand, if its system initially included a small inner planet, its initial eccentricity could have been several times its current value, and its internal heating an order of magnitude greater for the first quarter of its life.

The implication of these examples is that reliable inferences about the origin and evolution of a warm jupiter and its system are limited. Its initial eccentricity after formation is poorly constrained, as is the corresponding amount of early internal tidal heating. Moreover, the initial system may have contained more planets than currently observed. The tidal removal process may have been governed, or at least affected, the statistics of the multiplicity of planets in observed systems.

The examples here also show how we can place constraints on the properties of the initial inner planet (e.g. mass, semimajor axis, and $Q$) from the current survival of an inner planet.  In general, the smaller the mass of the inner planet, the more likely that it has survived to the present time (Figures \ref{HATmany} and \ref{HDmany}).  But before its demise even a smaller one could have had significant effects on its accompanying warm jupiter. In any case, the lifetime of an inner planet can be dramatically shorter than it would be without the presence of the warm jupiter.

For the specific cases considered here, tidal heating of the warm jupiters, while greatly enhanced by the inner planet's presence, would not be great enough to yield an anomalously large radius. Increasing the assumed size of the putative inner planet could enhance the heating further. On the other hand, a more massive inner planet would be less likely to be rocky, so its $Q$ would be larger.  Such a planet would be less likely to have fallen into the star and more likely to be currently observable.

Classical secular theory provides insight into what parameters of the inner planet would maximize effects on the warm Jupiter. For typical cases, including those examples in Sections \ref{infall} and \ref{damping}, one eigenmode of the secular interaction dies out so quickly that the evolution is governed by only the single remaining eigenmode. Tides damp the amplitude of that mode in proportion to the eccentricity values.  However, the actual damping of the eccentricities depends on the eigenvector, which describes how the mode is apportioned between the two planets.  The eigenvector depends on the semimajor axes and masses of the planets.  These considerations help constrain the system parameters that may most effectively allow secular interactions to enhance tidal effects (c.f. Sections \ref{deathheat} and \ref{dampingheat}).

For systems currently with two planets, like KOI-543, the parameters that control tidal evolution can be constrained by the requirement that initially, i.e. at the conclusion of the formation process, the planets had to be sufficiently separated for the system to have been stable, such considerations can be used to place limits on the current orbital eccentricities, given reasonable estimates of planetary $Q$. Such constraints are important for planets discovered by transits, where observations generally do not provide direct measurements of eccentricity, but planets are typically close enough to the star for tides to be important.

Although here we consider only examples of two-planet systems, similar types of the interactions and their effects on evolution and physical properties are to be expected in systems with three or more planets. For example, \citet{Ste13} have recently shown that, in general, very close-in planets appear to exist only if the next outward planet is not too close. That observation is consistent with our demonstration that an inner planet tends to be driven into its star if its eccentricity can be maintained by another planet. In any system where there are significant tides on or by at least one planet, considerable care is required in order to infer the range of possible past evolution scenarios and initial state of the system.

\acknowledgments
We thank Rory Barnes and Leslie Rogers for valuable discussions that helped this work. CVL was supported by a NASA Earth and Space Science Fellowship. This research has made use of the NASA Exoplanet Archive, which is operated by the California Institute of Technology, under contract with the National Aeronautics and Space Administration under the Exoplanet Exploration Program.


\clearpage


\appendix

\section{Secular Interactions with Tides in a Two-planet System} \label{appA}

Evolution of the rate of change of the amplitudes $E_m$ of the eigenmodes of a multi-planet system is straightforward if the change is due to a process that acts only to damp one or more of the planet's eccentricities. The damping rate $\dot{e}$ is generally proportional to $e$, so the secular equations (Eqns \ref{kdot}, \ref{hdot}) simply get additional linear terms. The eigenmode solution is obtained in the usual way, yielding complex eigenfrequencies whose real parts give the damping timescale for the amplitudes $E_m$.

However, tides act to change $a$ values along with $e$. As a result of the exchange of angular momentum among planets due to secular interactions, this change in $a$ also contributes to evolution of $E_m$ values. \citet{Gre11} derived formulae for this effect and added it to the change in $E_m$ due to damping of eccentricities.

Here we extend the results in the following ways (a) tides on both planets and tides on the star are included, (b) $\dot{a}$ and $\dot{e}$ are considered together rather than adding solutions due to each separately. The purpose is to have a way to evaluate the change over time of all the eccentricity components (all $e_{mp}$) by integrating only two $\dot{E}_m$, rather than four $\dot{e}_{mp}$, along with the two $\dot{a}_p$.

Also, the solution will enable further understanding of what parameters most strongly contribute to the ultimate eccentricity evolution. For example, the results derived here are applied  in Section \ref{infall} to reveal that the change in $e_1$ is due to a change in the relevant eigenvector, rather than a change in the eigenmode amplitude.

The derivation through Eqn \ref{dote22} (below) mimics that in \citet{Gre11} except that (a) the contributions from $\dot{e}_p$ and $\dot{a}_p$ are considered simultaneously, (b) tides on the star are included, and (c) tides on both planets are included. The equations are revisited here for completeness and homogeneity of notation. Then we derive the rates of change of the eigenmode amplitudes. We then derive Eqn \ref{reldamp}.

\begin{figure}
\epsscale{1.0}
\plotone{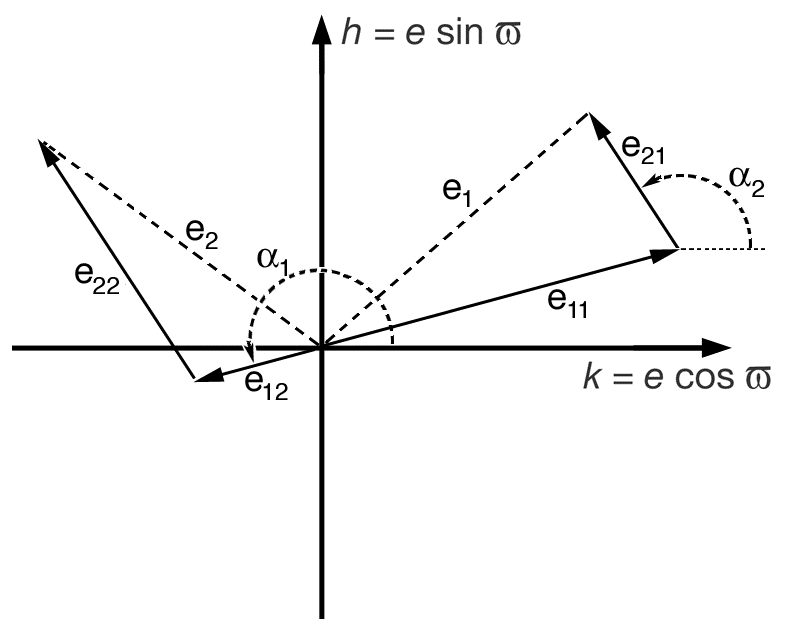}
\caption{Secular eigenmodes adding vectorally to planets' eccentricities in a two-planet system. Note that $e_{11}$ is negative, so it points $180^o$ from $\alpha_1$. Angles $\alpha_1$ and $\alpha_2$ rotate at rates given by the eigenfrequencies of the system. \label{ONE}}
\end{figure}

\begin{figure}
\epsscale{1.0}
\plotone{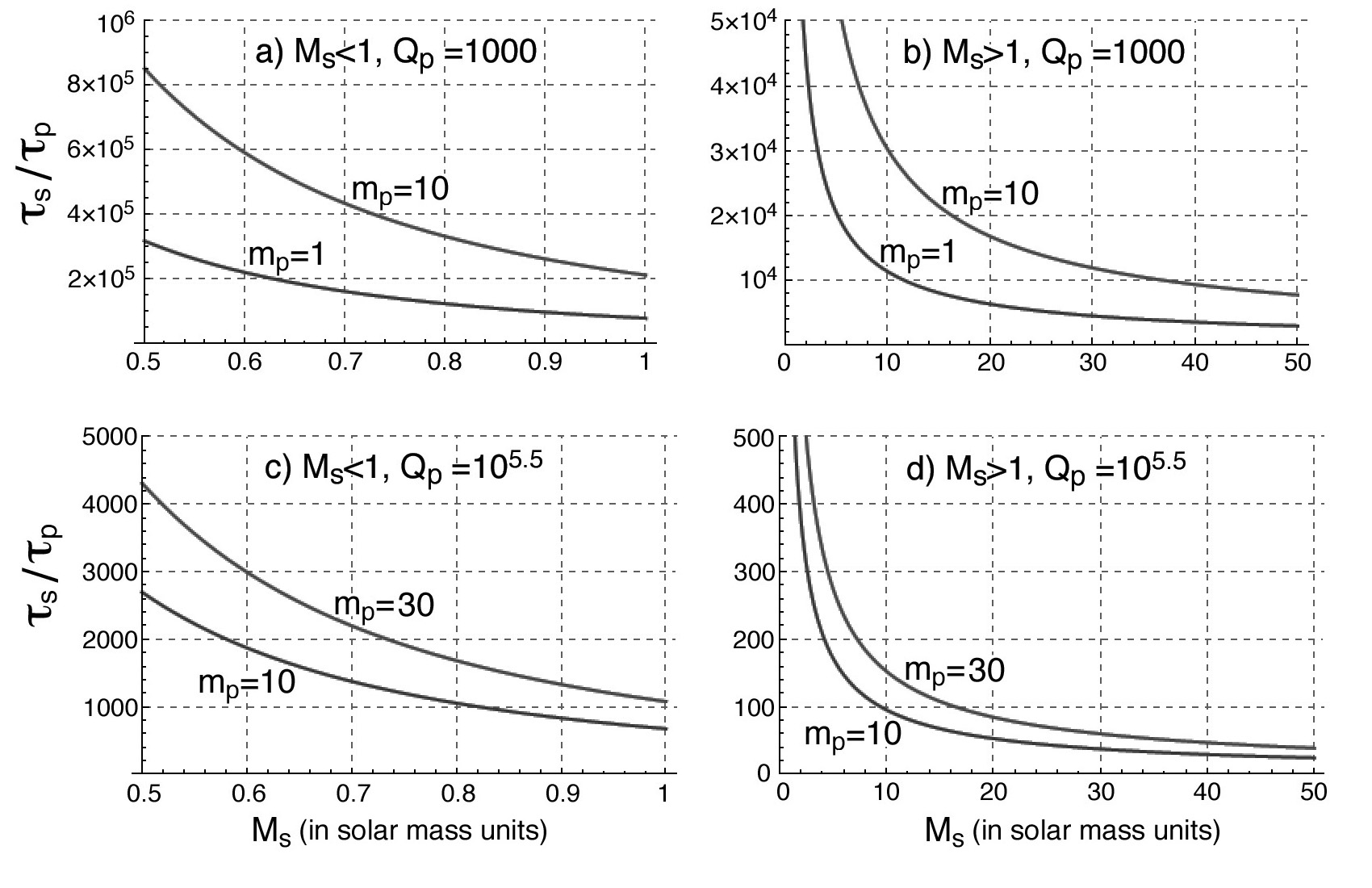}
\caption{Comparison of the rates of tidal evolution due to tides raised on the star on on the planet, given in terms of the ratio of evolution timescales $\tau_s/\tau_p$ (from Eqn \ref{tauratio}), where the subscript refers to the body on which the tide is raised. The top panels show the ratio as a function of stellar mass Ms for planets of mass 1 and 10 Earth masses and $Q=1000$. The bottom panels have planets of 10 and 30 Earth masses and $Q= 10^{5.5}$. Here we assume the planet's radius goes as $m_p^{1/2.06}$, and the stellar radius is proportional to $M_s^{0.8}$ for $M_s < M_{Solar}$ and $M_s^{0.57}$ for $M_s > M_{Solar}$. \label{plaVSste}}
\end{figure}

\begin{figure}
\epsscale{1.0}
\plotone{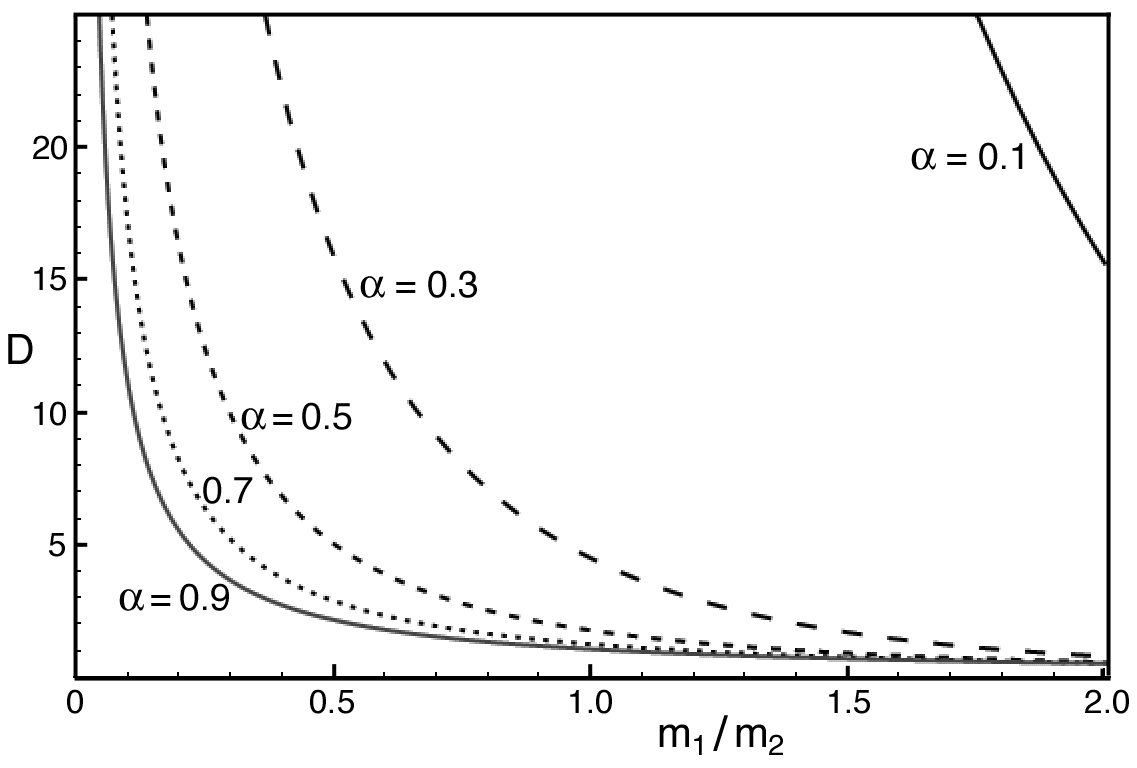}
\caption{The relative damping rates of the two eigenmode amplitudes for a system of two secularly interacting planets, where the inner planet's eccentricity is damped. The relative rates are expressed as $D\equiv (dot{E}_1/E_1)/(dot{E}_2/E_2)$, and shown as a function of semi-major axis ratio ($\alpha = a_1/a_2$) and mass ratio ($m_1/m_2$), as given by Eqn \ref{reldamp}. \label{DAMP1}}
\end{figure}

\begin{figure}
\epsscale{1.0}
\plotone{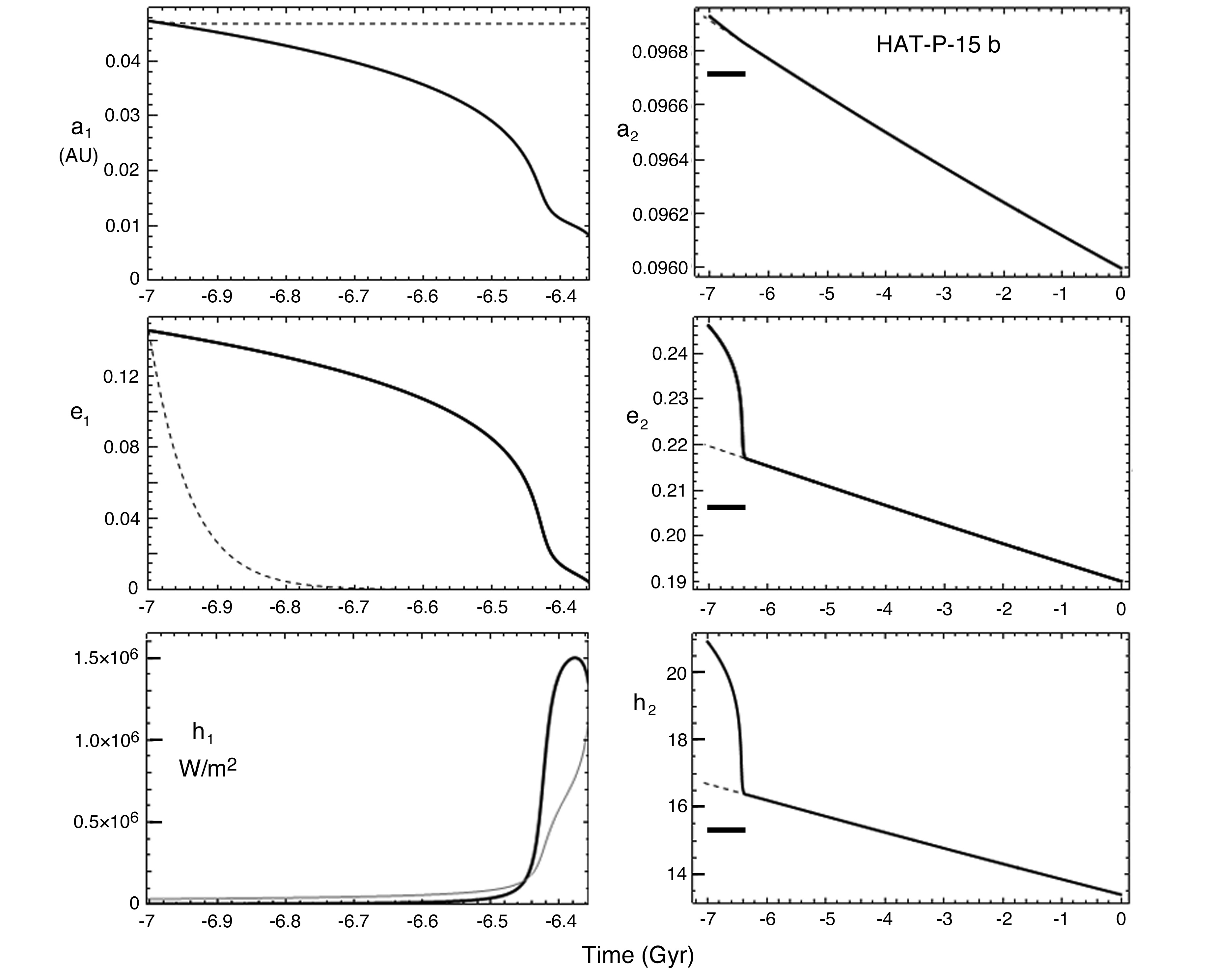}
\caption{Tidal evolution of the HAT-P-15 system with a 5 $M_{Earth}$ inner planet in addition to the known planet b. The left-hand panels show evolution of the inner planet, and the right-hand panels show the outer planet (HAT-P-15 b). Note the different time ranges. Top panels show semimajor axis as a function of time; middle panels show eccentricities; bottom panels show heat dissipation in each planet. The dashed lines show how each planet would evolve if it were the only one in the system. Solid lines show evolution of each planet, including secular interactions between the planets as well as tides on both planets and the star. The solid horizontal lines in the right hand panels denote the lifetime of the inner planet. In the bottom left panel, the solid gray line shows stellar insolation (“instellation”). Secular perturbations by the outer planet maintain the eccentricity of the inner one, so that the latter migrates inward reaching the star in only 0.65 Gyr.  Tidal heating exceeds the instellation.
\label{HATa}}
\end{figure}

\begin{figure}
\epsscale{0.7}
\plotone{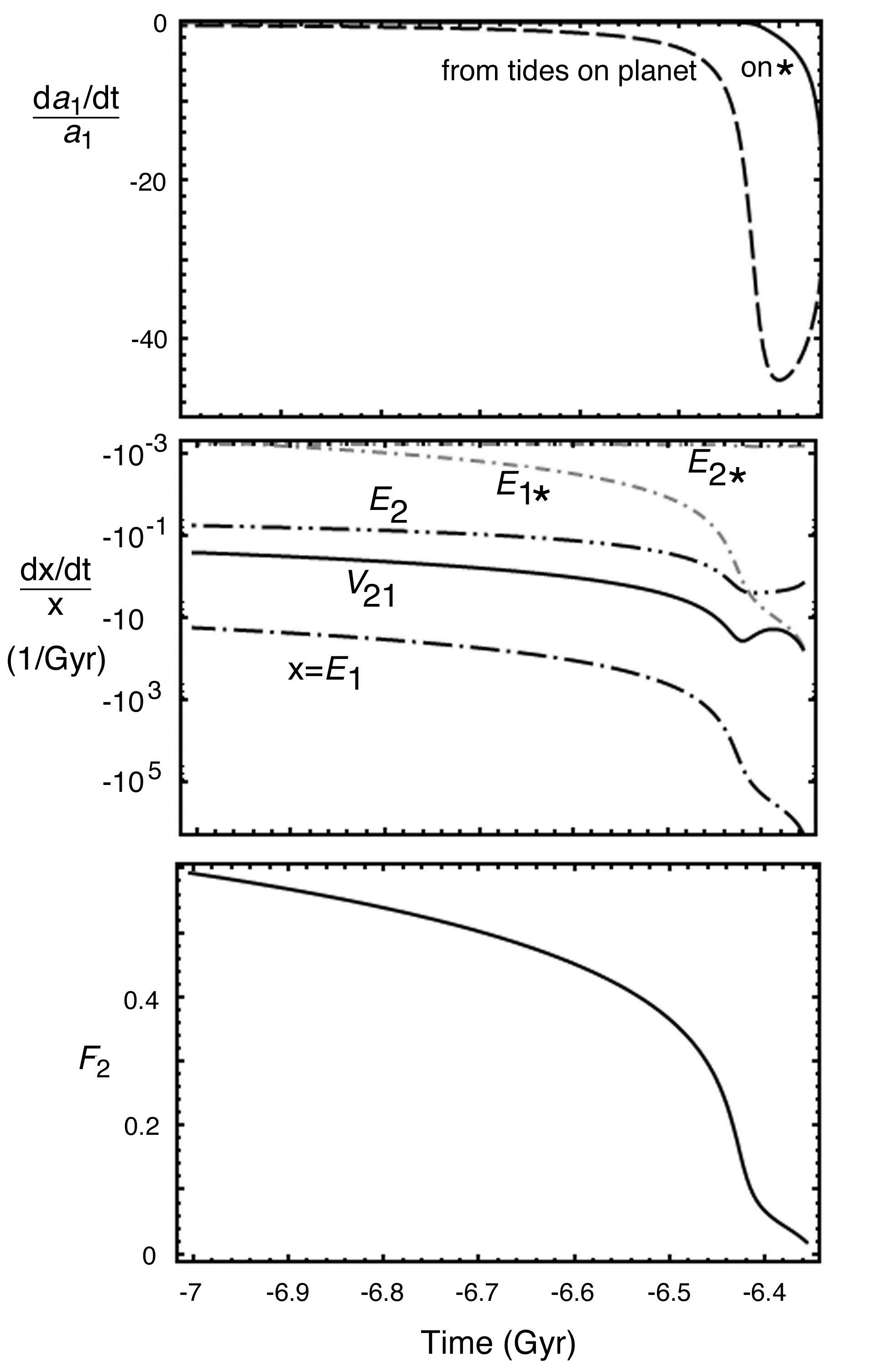}
\caption{Secular orbital parameters and their rates of change as functions of time for the system of HAT-P-15 b with the 5 $M_{Earth}$ inner companion. $Top$: The fractional rate of change of the inner planet's semimajor axis due to tides on the planets (dashed line) and due to tides on the star (solid line). $Middle$: The fractional rates of change of the eigenmode amplitudes $\dot{E}_1/E_1$ and $\dot{E}_2/E_2$ due to tides on the planets (black) and on the star (gray), as well as $\dot{V}_{21}/V_{21}$, the fractional rate of change of the eigenvector component $V_{21}$. $Bottom$: $F_2=V_{21}/V_{22}$, the ratio of eigenmode 2's contribution to the inner and outer planets respectively.
\label{HATb}}
\end{figure}

\begin{figure}
\epsscale{0.8}
\plotone{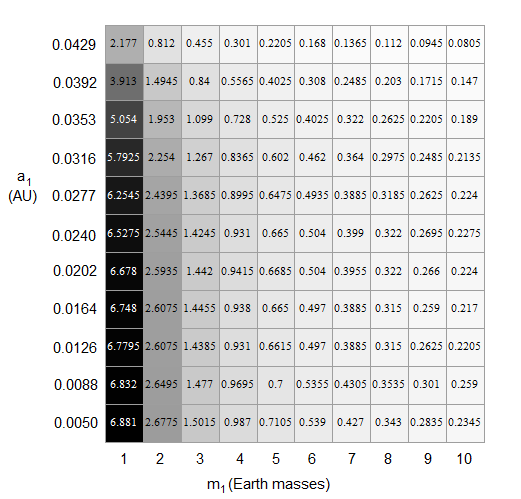}
\caption{For a hypothetical current inner companion to HAT-P-15 b with mass $m_1$ and semimajor axis $a_1$, the number in the box shows the time (in Gyr) since leaving either the 3:1 mean motion resonance or the limit of stability (whichever is shorter), assuming the inner planet's $Q=1000$. Given the estimated age of the system of about 7 Gyr, an inner planet of about one Earth mass could still survive in this system.
\label{HATmany}}
\end{figure}

\begin{figure}
\epsscale{1.0}
\plotone{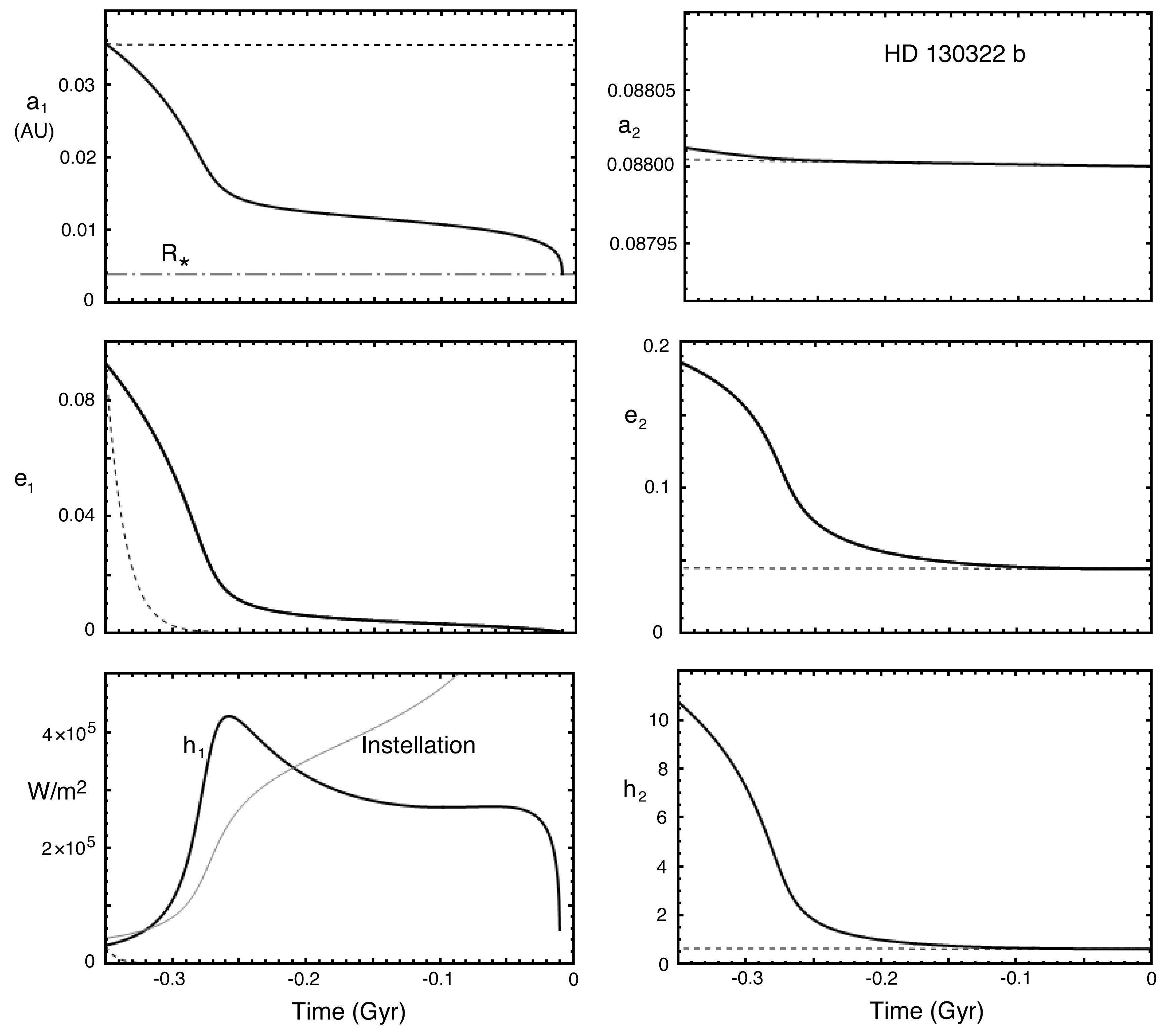}
\caption{HD 130322 with 10 $M_{Earth}$ companion c.f. Figure \ref{HATa}. The top right panel clearly shows that HD 130322 b's semi-major axis evolution does not differ from how it would evolve as a single planet. On the other hand HD 130322 b's eccentricity is strongly affected. 
\label{HD10a}}
\end{figure}

\begin{figure}
\epsscale{0.7}
\plotone{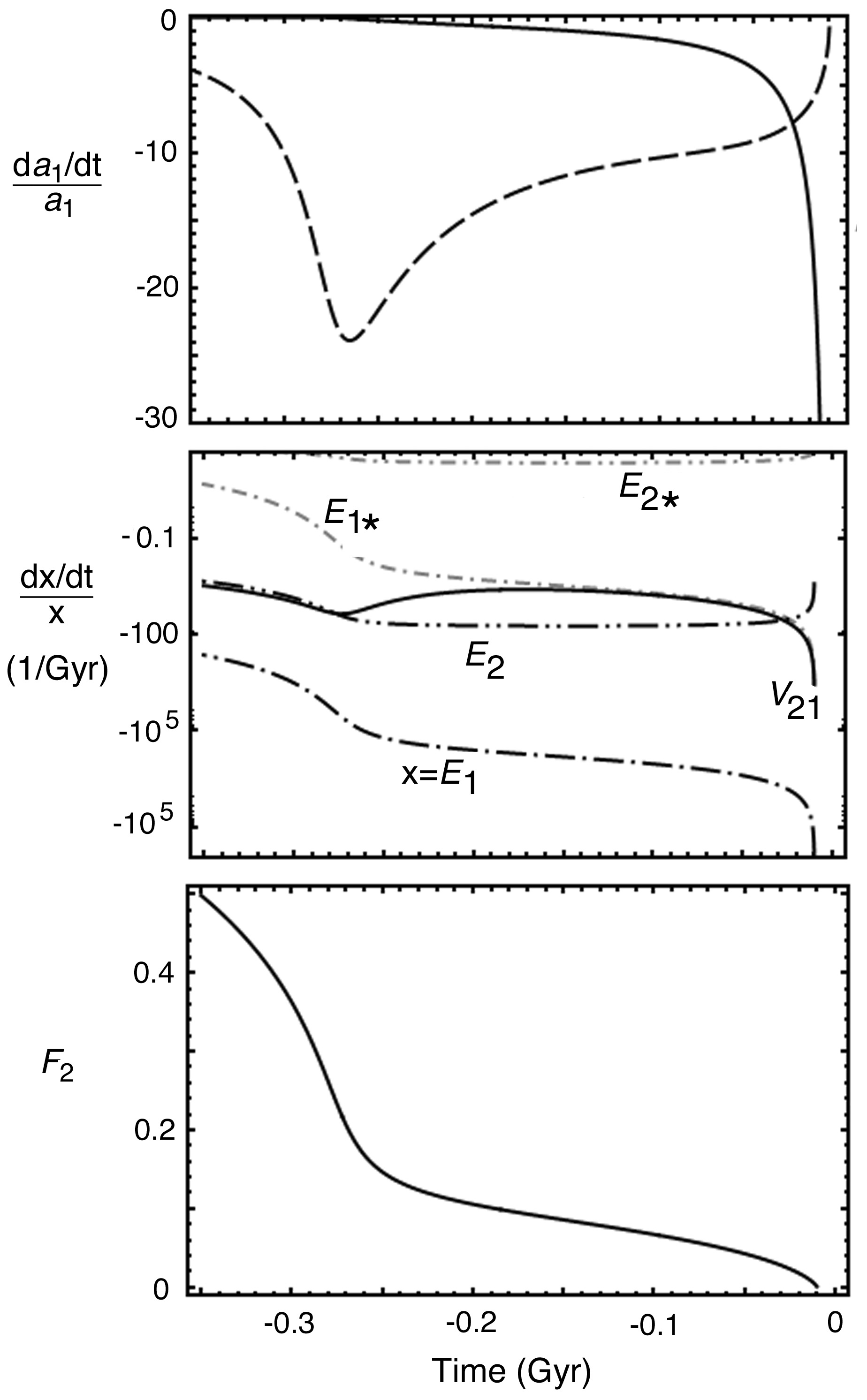}
\caption{Secular orbital parameters and their rates of change as functions of time for the system of HD 130322 with a hypothetical 10 $M_{Earth}$ inner planet (c.f. Fig. \ref{HATb}).
\label{HD10b}}
\end{figure}

\begin{figure}
\epsscale{0.8}
\plotone{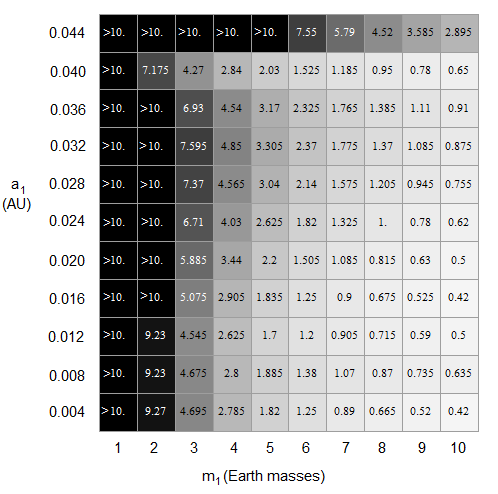}
\caption{For a hypothetical current inner companion to HD 130322 b with mass $m_1$ and semimajor axis $a_1$, the number in the box shows the time (in Gyr) since leaving either the 3:1 mean motion resonance or the limit of stability (whichever is shorter), assuming the inner planet's $Q=1000$ (c.f. Fig. \ref{HATmany}). Given the estimated age of the system of about 0.35 Gyr, any terrestrial scale inner planet with this $Q$ could still survive in this system. Note that the 3:1 resonance is located at approximately 0.042 AU.
\label{HDmany}}
\end{figure}

\begin{figure}
\epsscale{1.0}
\plotone{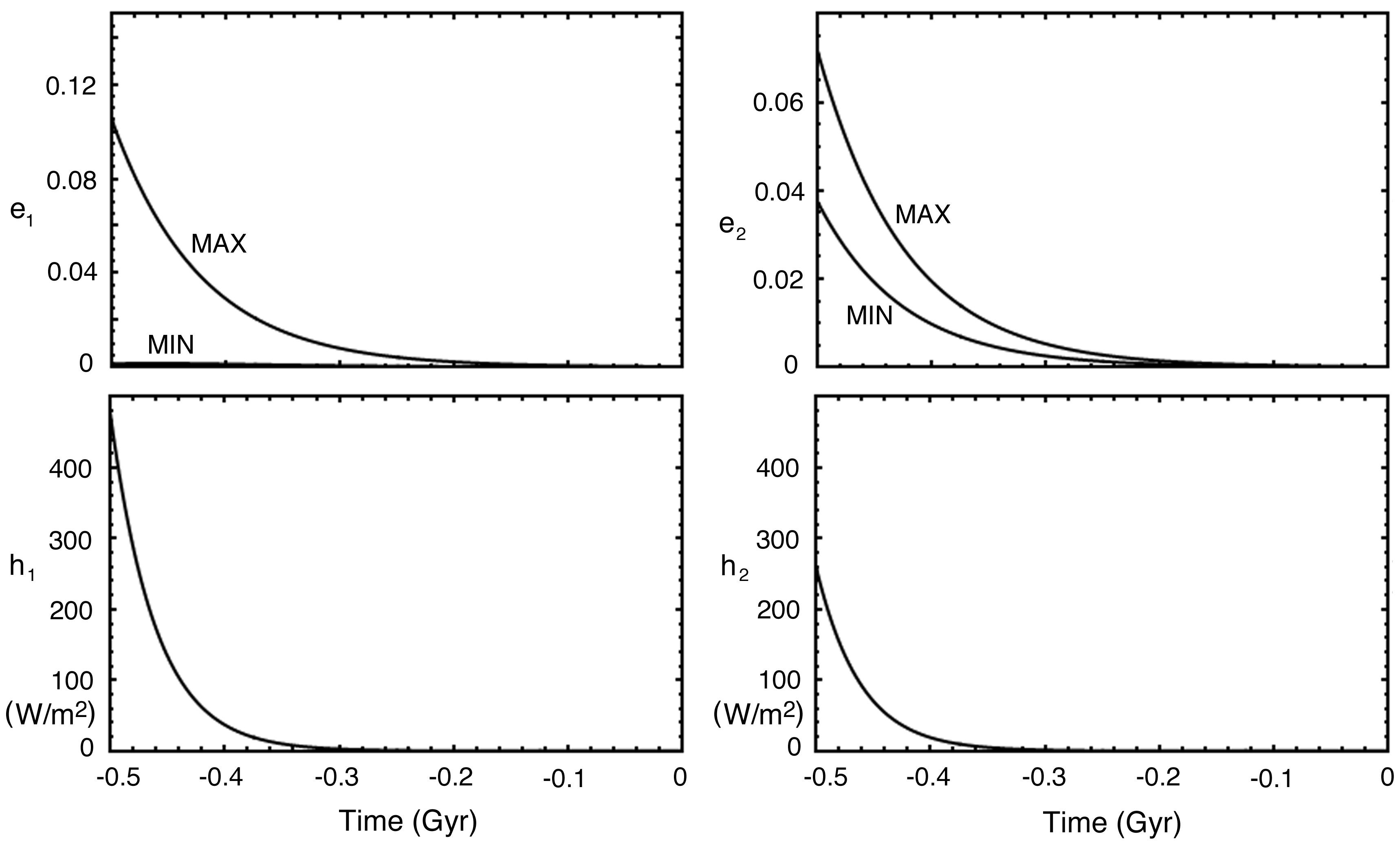}
\caption{Tidal evolution of the KOI 543 system, assuming $Q_p = 1000$ for both planets and current ($t = 0$) eigenmode amplitudes of $E_1 = E_2 = 10^{-4}$. In contrast to previous cases (e.g. Figs. \ref{HATa} and \ref{HD10a}), the modes have comparable amplitude during much of the evolution, so the eccentricities oscillate over the range shown due to the secular interactions. Because the damping timescales of both modes are similar, the amplitudes of the eccentricity oscillation change on a similar timescale to the mean eccentricities. The rate of tidal heating, shown in the lower panels, is significant, as discussed in Section \ref{constrain}.
\label{K5431000a}}
\end{figure}

\begin{figure}
\epsscale{0.65}
\plotone{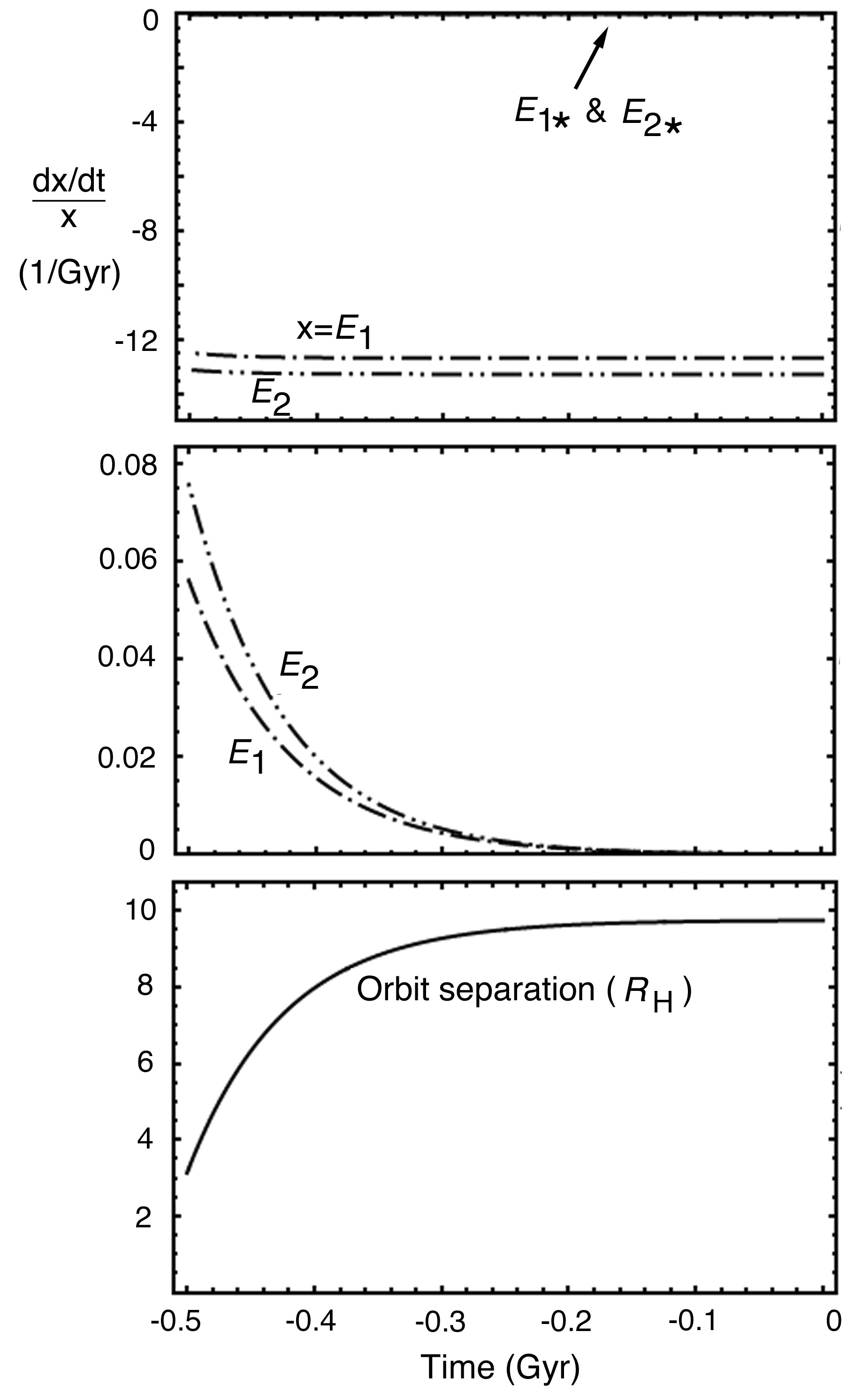}
\caption{Secular elements and rates of change for the KOI 543 system with $Q_p = 1000$ for both planets, during the same evolution displayed in Figure \ref{K5431000a}. $Top$: Similar to the middle panels of Figs. \ref{HATb} and \ref{HD10b}, showing the fractional rates of change of the eigenmode amplitudes $\dot{E}_1/E_1$ and $\dot{E}_2/E_2$ due to tides on the planets (black) and on the star (gray). $Middle$: The eigenmode amplitudes $E_1$ and $E_2$. $Bottom$: Minimum separation between the two planets (pericenter of the outer planet minus apocenter of the inner planet) in multiples of the Hill radius $R_H$. The system would have been only barely over the limit of stability (2.5 $R_{H}$) 0.5 Gyr ago, which constrains either $Q$ to $> 1000$, or $E_1$ and $E_2$ to $< 10^{-4}$ for this system.
\label{K5431000b}}
\end{figure}

\begin{figure}
\epsscale{1.0}
\plotone{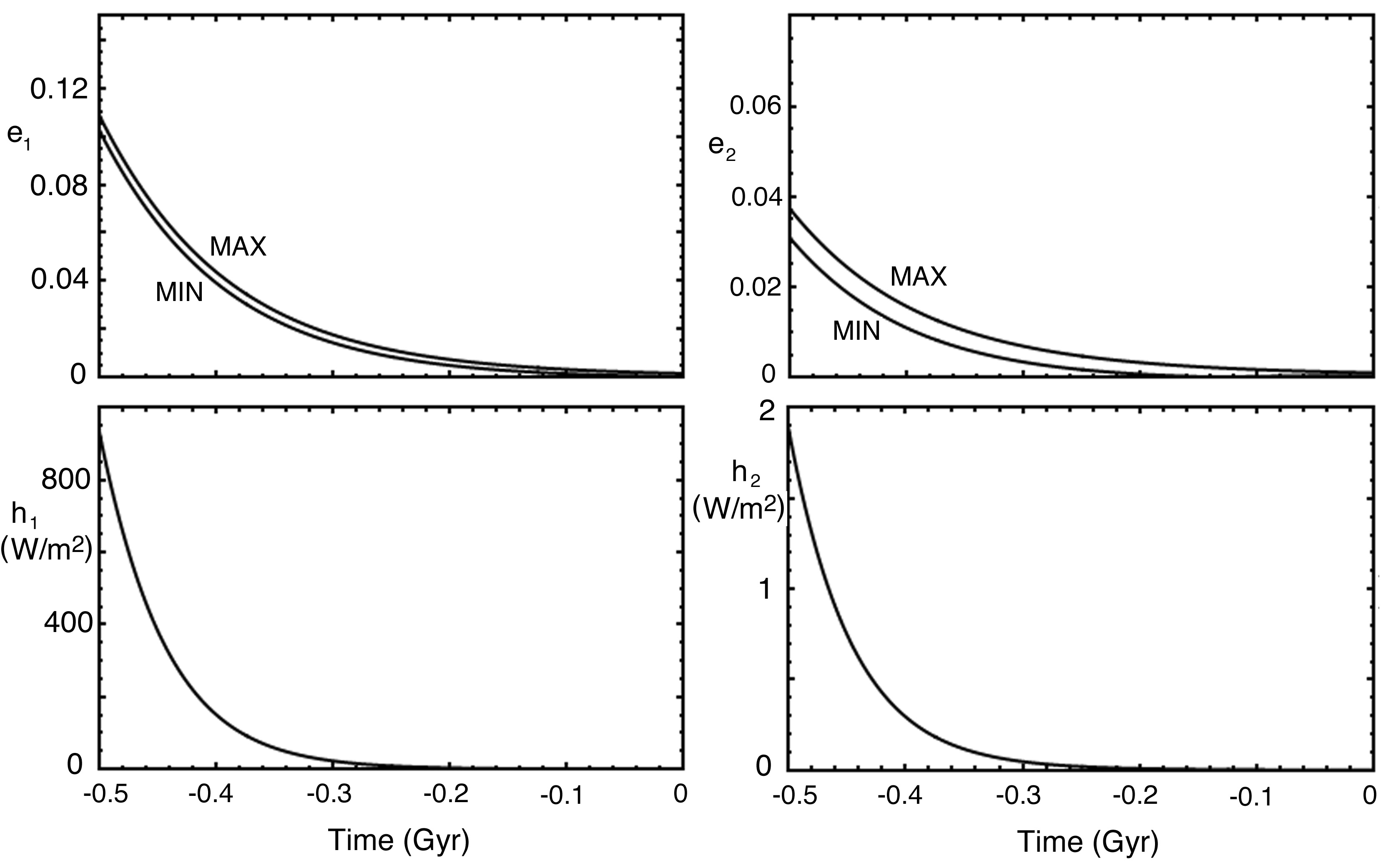}
\caption{Tidal evolution of the KOI 543 system, similar to Figure \ref{K5431000a} except here the outer planet's $Q$ is $5\times10^4$. In this case, the damping timescale for $E_1$ is noticeably faster than that of $E_2$ (see also Figure \ref{K543mixedb}). As a result the amplitudes of the eccentricity oscillation evolve more slowly than the mean eccentricities. The outer planet receives a much lower rate of tidal heating due to its high $Q$.
\label{K543mixeda}}
\end{figure}

\begin{figure}
\epsscale{0.6}
\plotone{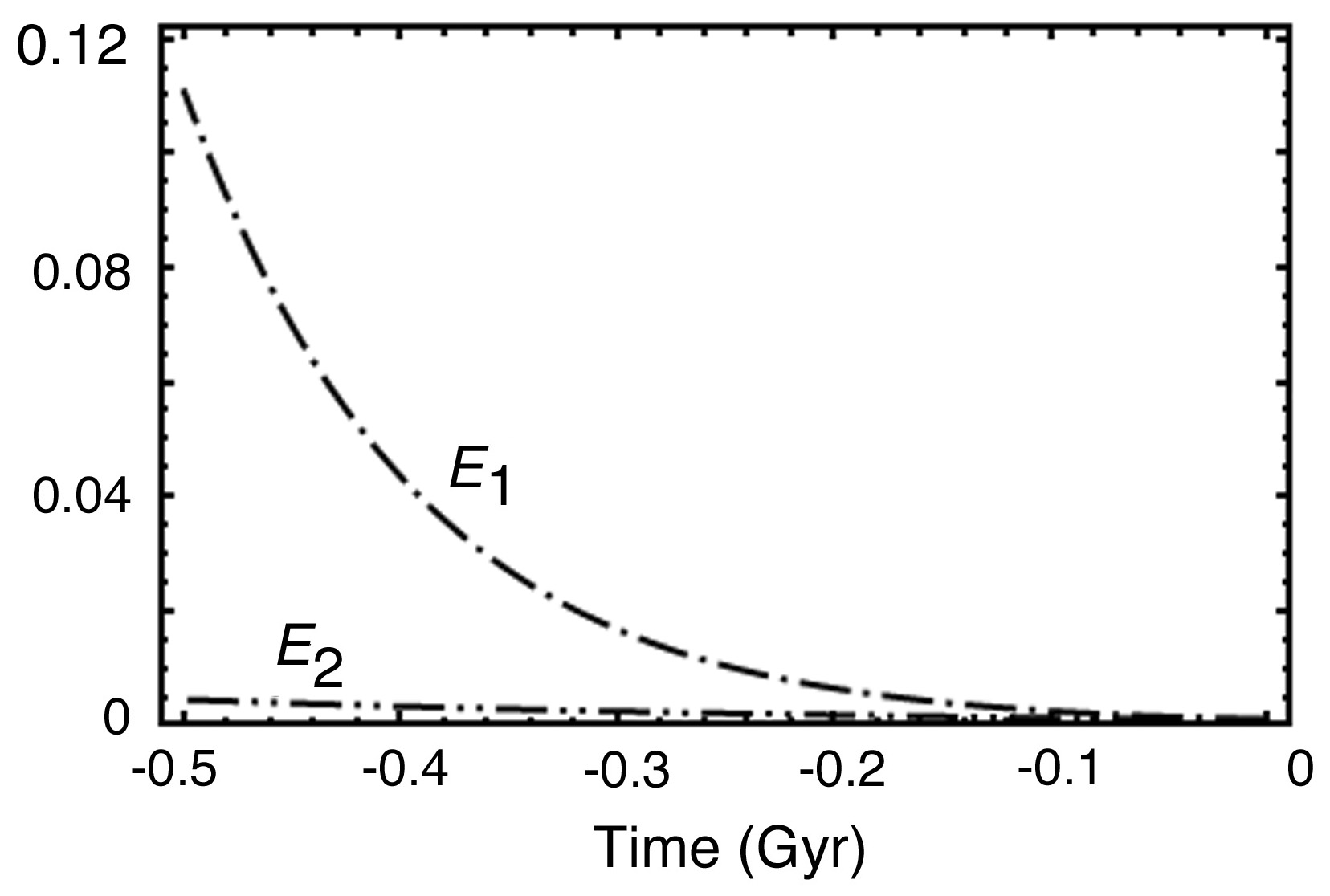}
\caption{Evolution of the eigenmode amplitudes $E_1$ (dash dot)  and $E_2$ (dash dot dot) for the KOI 543 system, for the same case shown in Figure 13 (c.f. middle panel of Figure \ref{K5431000b}).
KOI 543 with $Q_1=1000$ and $Q_2 = 5\times10^{4}$. Note here how $E_1$ evolves faster than $E_2$.
\label{K543mixedb}}
\end{figure}

For each planet, as shown in Figure \ref{ONE},

\begin{eqnarray}
k_p &=& e_p \cos \varpi_p \nonumber \\
&=& e_{1p} \cos \alpha_1 + e_{2p} \cos \alpha_2 \nonumber \\
&=& E_1 V_{1p} \cos \alpha_1 + E_2 V_{2p} \cos \alpha_2 \label{kp} \\
h_p &=& e_p \sin \varpi_p \nonumber \\
&=& e_{1p} \sin \alpha_1 + e_{2p} \sin \alpha_2 \nonumber \\
&=& E_1 V_{1p} \sin \alpha_1 + E_2 V_{2p} \sin \alpha_2 \label{hp}
\end{eqnarray}

\noindent where $\alpha_m = g_m t + \delta_m$ (Eqns \ref{kvst} and \ref{hvst}). In this The eccentricity of each planet is then

\begin{eqnarray}
e_p^2 &=& k_p^2 + h_p^2 \nonumber \\
&=& e_{1p}^2 + e_{2p}^2 + e_{1p} e_{2p} \cos\theta \label{epsq}
\end{eqnarray}

\noindent where $\theta = \alpha_1-\alpha_2$.

For a two-planet system, for each eigenmode, the ratio of its contribution to the inner planet ($p=1$) versus the outer planet ($p=2$) can be solved analytically:

\begin{equation} \label{F1}
F_1 \equiv \frac{V_{11}}{V_{12}}
= \frac{(A_{11}-A_{22})^2 +S}{2 A_{21}}
\end{equation}

\begin{equation} \label{F2}
F_2 \equiv \frac{V_{21}}{V_{22}}
= \frac{(A_{11}-A_{22})^2 -S}{2 A_{21}}
\end{equation}

\noindent where $S = \sqrt{ (A_{11}-A_{22})^2 + 4 A_{12} A_{21}}$ (e.g. Eqns 7 of \citet{Gre11}, with minor changes in notation). Note from Eqn \ref{emp} above that $e_{11}/e_{12}=F_1$ and $e_{21}/e_{22}=F_2$. For secular planet-planet interactions, and including other contributions to apsidal precession (e.g. GR), the $A_{ij}$ are functions of the stellar mass, the planetary masses, and the planet semi-major axes only \citep{Mur99}. If only planet-planet interactions are considered, then $F_1$ and $F_2$ depend only on the planetary mass and semi-major axis ratios ($m_1/m_2$ and $a_1/a_2$), because each $A_{ij}$ depends on the stellar mass in the same way. The eigenvectors are normalized ($V_{m1}^2 + V_{m2}^2 = 1$ for each eigenmode $m$). Thus, we can express the individual eigenvector components as

\begin{eqnarray}
V_{m1}= F_m / \sqrt{1+F_m^2} \label{Vm1} \\
V_m1= 1 / \sqrt{1+F_m^2} \label{Vm2}
\end{eqnarray}

\noindent (recall that the first subscript is the mode and the second is the planet).


\noindent Note that $F_1$ is negative, and thus $V_{11}$ is also negative.

We can evaluate how the secular state of the system will change under any given $\dot{a_p}$ and $\dot{e_p}$ by taking derivatives of $k_p$ (Eqn \ref{kp}) and $h_p$ (Eqn \ref{hp}).

\begin{eqnarray}
\delta k_1 = \delta e_{11} \cos\alpha_1 - \delta\alpha_1 e_{11} \sin\alpha_1 + \delta e_{21} \cos\alpha_2 - \delta\alpha_2 e_{21} \sin\alpha_2 \\
\delta h_1 = \delta e_{11} \sin\alpha_1 + \delta\alpha_1 e_{11} \cos\alpha_1 + \delta e_{21} \sin\alpha_2 + \delta\alpha_2 e_{21} \cos\alpha_2 \\
\delta k_2 = \delta e_{12} \cos\alpha_1 - \delta\alpha_1 e_{12} \sin\alpha_1 + \delta e_{22} \cos\alpha_2 - \delta\alpha_2 e_{22} \sin\alpha_2 \\
\delta h_2 = \delta e_{12} \sin\alpha_1 + \delta\alpha_1 e_{21} \cos\alpha_1 + \delta e_{22} \sin\alpha_2 + \delta\alpha_2 e_{22} \cos\alpha_2
\end{eqnarray}

The resulting equations can be combined to eliminate dependence on the individual $\alpha_m$ and $\delta \alpha_m$:

\begin{equation} \label{eqf1}
e_{22} \delta e_{11} - e_{21} \delta e_{12} + \cos\theta(e_{22} \delta e_{21} - e_{21} \delta e_{22} ) = f_1
\end{equation}

\noindent where

\begin{equation}
f_1 \equiv e_{22}(\delta k_1 \cos\alpha_1 + \delta h_1 \sin\alpha_1)
	- e_{21}(\delta k_2 \cos\alpha_1 + \delta h_2 \sin\alpha_1)
\end{equation}

\noindent and

\begin{equation} \label{eqf2}
e_{12} \delta e_{21} - e_{11} \delta e_{22} + \cos\theta(e_{12} \delta e_{11} - e_{11} \delta e_{12} ) = f_2
\end{equation}

\noindent where

\begin{equation}
f_2 \equiv e_{12}(\delta k_1 \cos\alpha_2 + \delta h_1 sin\alpha_2)
	- e_{11}(\delta k_2 \cos\alpha_2 + \delta h_2 \sin\alpha_2)
\end{equation}

We also know from Eqns \ref{kp} and \ref{hp} that

\begin{eqnarray}
\delta k_1 = \delta e_1 \cos\varpi_1 - \delta\varpi_1 e_1 \sin\varpi_1 \\
\delta h_1 = \delta e_1 \sin\varpi_1 + \delta\varpi_1 e_1 \cos\varpi_1 \\
\delta k_1 = \delta e_2 \cos\varpi_2 - \delta\varpi_2 e_2 \sin\varpi_2 \\
\delta k_1 = \delta e_2 \sin\varpi_2 + \delta\varpi_2 e_2 \cos\varpi_2
\end{eqnarray}

\noindent So if $\dot{e_1}/e_1 = T_1$ and $\dot{e_2}/e_2 = T_2$ (given by Eqn \ref{edot} for tides),

\begin{eqnarray}
f_1 = T_1 e_{22} (e_{11} + e_{21}\cos\theta) + T_2 e_{21} (e_{12} + e_{22}\cos\theta) \\
f_2 = T_1 e_{12} (e_{21} + e_{11}\cos\theta) + T_2 e_{11} (e_{22} + e_{12}\cos\theta)
\end{eqnarray}

Combining Eqns \ref{eqf1} and \ref{eqf2} with $\delta F_m = \delta e_{m1}/e_{m2} - \delta e_{m2} e_{m1}/e_{m2}^2$ (derivatives of Eqns \ref{F1} and \ref{F2}) gives us four equations with four unknown $\delta e_{mp}$. Solving for the $\delta e_{mp}$ yields:

\begin{eqnarray}
\delta e_{11} &=& \frac{1}{e_{12}e_{21}-e_{11}e_{22}} \left(
e_{12}^2 e_{21} \delta F_1 +  e_{11} e_{22}^2 \delta F_2 \cos\theta - e_{11} f_1 \right) \label{dele11} \\
\delta e_{12} &=& \frac{1}{e_{12}e_{21}-e_{11}e_{22}} \left(
e_{12}^2 e_{22} \delta F_1 + e_{12} e_{22}^2 \delta F_2 \cos\theta  - e_{12} f_1 \right) \\
\delta e_{21} &=& \frac{-1}{e_{12}e_{21}-e_{11}e_{22}} \left(
e_{12}^2 e_{21} \delta F_1 \cos\theta  + e_{11} e_{22}^2 \delta F_2 - e_{21} f_2 \right) \\
\delta e_{22} &=& \frac{-1}{e_{12}e_{21}-e_{11}e_{22}} \left(
e_{12}^2 e_{22} \delta F_1 \cos\theta + e_{12} e_{22}^2 \delta F_2 - e_{22} f_2 \right) \label{dele22}
\end{eqnarray}

\noindent $\delta F_m$ describe how the eigenvectors change due to changing masses or semi-major axes. That is, if the eigenvectors are changing due to changing semi-major axes then $\delta{F}_m = (dF_m/da_1)\dot{a}_1 + (dF_m/da_2)\delta{a}_2$. It should be noted that while we are focusing our discussion on tides and thus changes imposed by one or more $\dot{a_p}$, these equations could also be used to calculate the changes in the eigenmodes due to change in planetary masses $\dot{m_p}$.

Consider $\dot{a}_p$ of the form:

\begin{equation}
\dot{a}_p = G_p + H_p e_p^2 \label{adotform}
\end{equation}

\noindent For tides, planetary tides are contained in $H_p$, and tides on the star appear in both $G_p$ and $H_p$ (compare to Eqn \ref{adot}). Inserting Eqn \ref{adotform} into Eqns \ref{dele11} to \ref{dele22} yields:

\begin{eqnarray}
\dot{e}_{11} = \frac{1}{e_{12}e_{21}-e_{11}e_{22}} &\Bigg(&
e_{12}^2 e_{21} \sum_p \left[ \frac{\partial F_1}{\partial a_p} (G_p + H_p e_p^2) \right] \\
&+&  e_{11} e_{22}^2 \cos\theta \sum_p \left[ \frac{\partial F_2}{\partial a_p} (G_p + H_p e_p^2) \right]
- e_{11} f_1 \Bigg) \\
\dot{e}_{12} = \frac{1}{e_{12}e_{21}-e_{11}e_{22}} &\Bigg(&
e_{12}^2 e_{22} \sum_p \left[ \frac{\partial F_1}{\partial a_p} (G_p + H_p e_p^2) \right] \\
&+& e_{12} e_{22}^2 \cos\theta \sum_p \left[ \frac{\partial F_2}{\partial a_p} (G_p + H_p e_p^2) \right]
- e_{12} f_1 \Bigg) \\
\dot{e}_{21} = \frac{-1}{e_{12}e_{21}-e_{11}e_{22}} &\Bigg(&
e_{12}^2 e_{21} \cos\theta \sum_p \left[ \frac{\partial F_1}{\partial a_p} (G_p + H_p e_p^2) \right] \\
&+& e_{11} e_{22}^2 \sum_p \left[ \frac{\partial F_2}{\partial a_p} (G_p + H_p e_p^2) \right]
- e_{21} f_2 \Bigg) \\
\dot{e}_{22} = \frac{-1}{e_{12}e_{21}-e_{11}e_{22}} &\Bigg(&
e_{12}^2 e_{22} \cos\theta \sum_p \left[ \frac{\partial F_1}{\partial a_p} (G_p + H_p e_p^2) \right] \\
&+& e_{12} e_{22}^2 \sum_p \left[ \frac{\partial F_2}{\partial a_p} (G_p + H_p e_p^2) \right]
- e_{22} f_2 \Bigg) 
\end{eqnarray}

\noindent If tides (or whatever is causing the $\dot{a}_p$ and $\dot{e}_p$) happens on timescales that are long compared to the secular timescale then we can average over the secular cycle, i.e. average over $\theta$ ($=\alpha_1-\alpha_2$). In doing this we need to remember that $e_1^2$ and $e_2^2$ have terms that are independent of $\theta$ and those that depend on $\cos\theta$, Eqn \ref{epsq}. This averaging yields:

\begin{eqnarray}
\dot{e}_{11} = \frac{1}{e_{12}e_{21}-e_{11}e_{22}} &\Bigg(&
e_{12}^2 e_{21} \sum_p \left[ \frac{\partial F_1}{\partial a_p} (G_p + H_p (e_{1p}^2+e_{2p}^2)) \right] \\
&+&  e_{11} e_{22}^2 \sum_p \left[ \frac{\partial F_2}{\partial a_p} \frac{H_p e_{1p} e_{2p}}{2} \right] \nonumber \\
&-& e_{11} (e_{11} e_{22} T_1 + e_{12} e_{21} T_2) \Bigg) \nonumber \\
\dot{e}_{12} = \frac{1}{e_{12}e_{21}-e_{11}e_{22}} &\Bigg(&
e_{12}^2 e_{22} \sum_p \left[ \frac{\partial F_1}{\partial a_p} (G_p + H_p (e_{1p}^2+e_{2p}^2)) \right] \\
&+& e_{12} e_{22}^2 \sum_p \left[ \frac{\partial F_2}{\partial a_p} \frac{H_p e_{1p} e_{2p}}{2} \right] \nonumber \\
&-& e_{12} (e_{11} e_{22} T_1 + e_{11} e_{21} T_2) \Bigg) \nonumber \\
\dot{e}_{21} = \frac{-1}{e_{12}e_{21}-e_{12}e_{22}} &\Bigg(&
e_{12}^2 e_{21} \sum_p \left[ \frac{\partial F_2}{\partial a_p} \frac{H_p e_{1p} e_{2p}}{2} \right] \\
&+& e_{11} e_{22}^2 \sum_p \left[ \frac{\partial F_1}{\partial a_p} (G_p + H_p (e_{1p}^2+e_{2p}^2)) \right] \nonumber \\
&-& e_{21} (e_{12} e_{21} T_1 + e_{11} e_{22} T_2) \Bigg) \nonumber \\
\dot{e}_{22} = \frac{-1}{e_{12}e_{21}-e_{11}e_{22}} &\Bigg(&
e_{12}^2 e_{22} \sum_p \left[ \frac{\partial F_2}{\partial a_p} \frac{H_p e_{1p} e_{2p}}{2} \right] \label{dote22} \\
&+& e_{12} e_{22}^2 \sum_p \left[ \frac{\partial F_1}{\partial a_p} (G_p + H_p (e_{1p}^2+e_{2p}^2)) \right] \nonumber \\
&-& e_{22} (e_{12} e_{21} T_1 + e_{11} e_{22} T_2) \Bigg) \nonumber
\end{eqnarray}

\noindent So, each $\dot{e}_{mp}$ is affected by changes in the secular structure (ultimately the semi-major axes) and eccentricity damping.

Each $e_{mp}$ is the product of eigenmode amplitude $E_m$ and eigenvector component $V_{mp}$. The derivatives of the $V_{mp}$ are obtained by differentiating Eqns \ref{Vm1} and \ref{Vm2}:

\begin{eqnarray}
\dot{V}_{11} &=& \frac{1}{(1+F_1^2)^{3/2}} \dot{F}_1 \label{dV11} \\
&=& \frac{1}{(1+F_1^2)^{3/2}} \sum_p \left[ \frac{\partial F_1}{\partial a_p} (G_p + H_p e_p^2) \right] \nonumber \\
\dot{V}_{12} &=& \frac{-F_1}{(1+F_1^2)^{3/2}} \dot{F}_1 \\
&=& \frac{-F_1}{(1+F_1^2)^{3/2}} \sum_p \left[ \frac{\partial F_1}{\partial a_p} (G_p + H_p e_p^2) \right] \nonumber \\
\dot{V}_{21} &=& \frac{1}{(1+F_2^2)^{3/2}} \dot{F}_2 \\
&=& \frac{1}{(1+F_2^2)^{3/2}} \sum_p \left[ \frac{\partial F_2}{\partial a_p} (G_p + H_p e_p^2) \right] \nonumber \\
\dot{V}_{22} &=& \frac{-F_2}{(1+F_2^2)^{3/2}} \dot{F}_2 \label{dV22} \\
&=& \frac{-F_1}{(1+F_2^2)^{3/2}} \sum_p \left[ \frac{\partial F_2}{\partial a_p} (G_p + H_p e_p^2) \right] \nonumber
\end{eqnarray}

Averaging over $\theta$ yields:

\begin{eqnarray}
\dot{V}_{11} = \frac{1}{(1+F_1^2)^{3/2}} \sum_p \left[ \frac{\partial F_1}{\partial a_p} (G_p + H_p (e_{1p}^2+e_{2p}^2)) \right] \\
\dot{V}_{12} = \frac{-F_1}{(1+F_1^2)^{3/2}} \sum_p \left[ \frac{\partial F_1}{\partial a_p} (G_p + H_p (e_{1p}^2+e_{2p}^2)) \right] \\
\dot{V}_{21} = \frac{1}{(1+F_2^2)^{3/2}} \sum_p \left[ \frac{\partial F_2}{\partial a_p} (G_p + H_p (e_{1p}^2+e_{2p}^2)) \right] \\
\dot{V}_{22} = \frac{-F_2}{(1+F_2^2)^{3/2}} \sum_p \left[ \frac{\partial F_2}{\partial a_p} (G_p + H_p (e_{1p}^2+e_{2p}^2)) \right]
\end{eqnarray}

We can derive what the change in eigenmode amplitude is by differentiating $e_{mp} = E_m V_{mp}$:

\begin{equation}
\dot{e}_{mp} = \dot{E}_m V_{mp} + E_m \dot{V}_{mp}
\end{equation}

\noindent and rearranging:

\begin{eqnarray}
\dot{E_1} &=& \frac{1}{V_{11}} ( \dot{e}_{11} - E_1 \dot{V_{11}} ) \nonumber \\
&=& \frac{1}{V_{12}} ( \dot{e}_{12} - E_1 \dot{V_{12}} ) \label{dE1} \\
\dot{E_2} &=& \frac{1}{V_{21}} ( \dot{e}_{21} - E_2 \dot{V_{21}} ) \nonumber \\
&=& \frac{1}{V_{22}} ( \dot{e}_{22} - E_2 \dot{V_{22}} ) \label{dE2}
\end{eqnarray}

\noindent Each $\dot{E}_m$ can be calculated from either $\dot{e}_{m1}$ or $\dot{e}_{m2}$. (In our numerical integrations we calculate each $\dot{E}_m$ both ways as a check for errors.) Note that it is possible for the eigenmode amplitudes to increase. Also note that changing the underlying eigenvectors can change both the eigenmode amplitudes (Eqns \ref{dE1} and \ref{dE2}), but eccentricity damping can only change the eigenmode amplitudes (Eqns \ref{dE1} and \ref{dE2}, and \ref{dV11} to \ref{dV22}).

If the change in the eigenmode amplitudes due to semi-major axis change are negligible, i.e. if $\dot{E_1} \approx \dot{e_{11}}/V_{11} \approx E_1 f_1$ and $\dot{E_2} \approx E_2 f_2$, and we average over $\theta$ then

\begin{eqnarray}
\frac{\dot{E_1}}{E_1} = \frac{-1}{V_{12}V_{21}-V_{11}V_{22}} (
-V_{11}V_{22} \frac{\dot{e}_1}{e_1} + V_{12}V_{21} \frac{\dot{e}_2}{e_2} ) \\
\frac{\dot{E_2}}{E_2} = \frac{-1}{V_{12}V_{21}-V_{11}V_{22}} (
-V_{12}V_{12} \frac{\dot{e}_1}{e_1} + V_{11}V_{22} \frac{\dot{e}_2}{e_2} )
\end{eqnarray}

\noindent (Remember that we have negative $V_{11}$.) If tides on the outer planet are negligible then $\dot{e}_2 \approx 0$ and thus

\begin{equation}
\frac{\dot{E}_1 / E_1}{\dot{E}_2 / E_2} = \frac{- V_{11} V_{22}}{V_{12} V_{21}}
\end{equation}

\noindent which is Eqn \ref{reldamp} in the main text.

\clearpage


\clearpage

\clearpage

\clearpage

\clearpage

\clearpage

\clearpage

\clearpage

\clearpage

\clearpage

\clearpage

\clearpage

\clearpage

\clearpage

\end{document}